%% file: main.tex
\documentclass[floatfix,aps,prd,nofootinbib,superscriptaddress,preprint]{revtex4}
\pdfoutput=1
\usepackage{amsmath,amssymb}
\usepackage{bm,bbm}
\usepackage{graphicx, graphics, color}
\usepackage{slashed}
\usepackage{booktabs}
\usepackage{tabu}
\usepackage[dvipsnames]{xcolor}
\usepackage[normalem]{ulem}
\newcommand{\fref}[1]{Fig.~\ref{f.#1}}

\newcommand{\eref}[1]{Eq.~(\ref{e.#1})}

\newcommand{\beq}{\begin{eqnarray}}
\newcommand{\eeq}{\end{eqnarray}}
\newcommand{\beqs}{\begin{eqnarray*}}
\newcommand{\eeqs}{\end{eqnarray*}}

\usepackage{tikz}
\usetikzlibrary{shapes,arrows}
\usetikzlibrary{positioning}
\newcommand{\chis}{\chi^2}
\newcommand{\chisdof}{\chi^2_{\rm dof}}

\newcommand{\mell}[1]{{\bm #1}}

\begin{document}

\title{\mbox{Iterative Monte Carlo analysis of spin-dependent
	parton distributions}}
\author{Nobuo Sato}
\affiliation{Jefferson Lab, Newport News, Virginia 23606, USA}
\author{W. Melnitchouk}
\affiliation{Jefferson Lab, Newport News, Virginia 23606, USA}
\author{S. E. Kuhn}
\affiliation{Old Dominion University, Norfolk, Virginia 23529, USA}
\author{J. J. Ethier}
\affiliation{College of William and Mary, Williamsburg, Virginia 23187, USA}
\author{A. Accardi}
\affiliation{Hampton University, Hampton, Virginia 23668, USA \\
\vspace*{0.2cm}
{\bf Jefferson Lab Angular Momentum (JAM) Collaboration
\vspace*{0.2cm} }}
\affiliation{Jefferson Lab, Newport News, Virginia 23606, USA}

\begin{abstract}
We present a comprehensive new global QCD analysis of polarized
inclusive deep-inelastic scattering, including the latest
high-precision data on longitudinal and transverse polarization
asymmetries from Jefferson Lab and elsewhere.
The analysis is performed using a new iterative Monte Carlo fitting
technique which generates stable fits to polarized parton distribution
functions (PDFs) with statistically rigorous uncertainties.
Inclusion of the Jefferson Lab data leads to a reduction in the PDF
errors for the valence and sea quarks, as well as in the gluon
polarization uncertainty at $x \gtrsim 0.1$.
The study also provides the first determination of the
flavor-separated twist-3 PDFs and the $d_2$ moment of
the nucleon within a global PDF analysis.
\end{abstract}

\date{\today}
\maketitle

\section{Introduction}
\label{s.intro}

The last few years have witnessed tremendous progress in our
understanding of the basic decomposition of the proton's spin
into its quark and gluon constituent parts, both in terms of
moments of spin-dependent parton distribution functions (PDFs)
and in their dependence on the momentum fraction $x$ carried by
the individual partons \cite{Lampe98, Aidala12, PJD13, Leader14}.
Recent data on inclusive jet \cite{STAR_jet15} and pion
\cite{PHENIX_pi14, PHENIX_pi15} production in polarized $pp$
collisions at the Relativistic Heavy Ion Collider (RHIC),
as well as double spin asymmetries from open charm muonproduction
at COMPASS \cite{COMPASS_g13}, have led to significant improvement
in the determination of the polarized gluon distribution at
small $x$ \cite{DSSV14}.
New results on longitudinal single-spin asymmetries in $W^\pm$
boson production \cite{STAR_W14, PHENIX_W15} are also yielding
better constraints on the polarization of sea quarks and antiquarks.

In fixed-target deep-inelastic scattering (DIS) experiments,
new high-precision data from Jefferson Lab on polarized protons
\cite{eg1a, eg1b-p-Prok, eg1b-p, eg1-dvcs}, deuterons \cite{eg1-dvcs,
eg1b-d} and $^3$He nuclei \cite{E06-014_A1, E06-014_d2, E01-012}
are yielding a wealth of information on nucleon spin structure
at lower energies.  As well as improving the constraints on the
large-$x$ behavior of polarized PDFs, the new results are also
providing new insights into nonperturbative quark-gluon
interaction effects through higher twist contributions.

In a previous study \cite{JAM13}, the Jefferson Lab Angular Momentum
(JAM) Collaboration performed a first analysis of inclusive longitudinal
and transverse polarization data down to low values of four-momentum
transfer squared $Q^2$ (= 1~GeV$^2$) and hadronic final state masses
squared $W^2$ (= 3.5~GeV$^2$), systematically taking into account
finite-$Q^2$ and nuclear corrections that are necessary at these
kinematics.  The increased statistics afforded by the weaker cuts
--- almost doubling the number of DIS data points --- resulted in more
reliable determinations of PDFs, particularly at large values of $x$.
In order to avoid dealing with the complications associated with
higher twist and nuclear corrections, many PDF analyses impose more
stringent cuts on $Q^2$ and $W^2$, which unfortunately eliminates
much of the data at the highest $x$ values.

Most of the existing phenomenological spin-dependent PDF analyses
\cite{DSSV09, LSS10, BB10, AAC09, AKS13} also utilize standard PDF
fitting technology, in which single fits are performed assuming a
basic parametric form for the PDFs, with the parameters obtained by
minimizing the overall $\chi^2$ of the fit.  The PDF errors are then
typically computed using the Hessian or Lagrange multiplier methods.
A drawback of this approach is that some of the shape parameters do
not play a significant role in describing the data, and attempts to
fix their values can be rather arbitrary due to correlations among
the distributions.  In some cases this can lead to overfitting,
with the $\chi^2$ per degree of freedom $\chi^2_{\rm dof} \ll 1$.
Furthermore, since the $\chi^2$ is a highly nonlinear function of
the fit parameters, in general there will be many solutions and
multiple local minima.
In practice, the extensive experience gained over the past two
decades with global QCD analyis of leading twist PDFs can be
exploited to render relatively stable results through judicious
choices for the starting parameters in the $\chi^2$ minimization.
One can also tune the number of free parameters in the fits to
reduce the number of solutions, even though the solutions can
never be guaranteed to be unique.  On the other hand, very limited
experience exists in fitting parameters for higher twist distributions
\cite{JAM13, ABMS09, BB12}, for which the signals are generally
smaller and the kinematic window for maximizing the sensitivity
of the fits to their presence is significantly narrower.

Because of these complications, in this work we propose an alternative
approach to global PDF analysis, based on a new iterative Monte Carlo
(IMC) fitting technique that allows a more robust extraction of both
leading and higher twist PDFs, with statistically rigorous PDF
uncertainties.
The idea behind this new iterative approach is to systematically
transform the priors obtained initially from a flat Monte Carlo
sampling into posteriors that are distributed consistently with
the information contained within the data.  Our method shares some
similarities with other Monte Carlo approaches, such as that by the
NNPDF group \cite{NNPDF14}, who also employ data resampling techniques
but use neural networks instead of traditional parametrizations.  In
particular, we retain the basic parametric form used in standard PDF
fitting, but maximally explore the parameter space using Monte Carlo
sampling, together with data resampling and cross-validation of the fit.
This avoids systematic biases introduced by performing single fits based
on an initial guess of the starting PDF parameters, and obviates the
necessity of fixing parameters that are not well constrained by data.

To offset the additional expense associated with performing
thousands of fits in the IMC approach, we perform all our
calculations in Mellin space, in analogy to the methodology
adopted by the DSSV group \cite{DSSV14, DSSV09}.
This requires the implementation of fast evaluation of nuclear
smearing \cite{KM08, KM09, Ethier13} and target mass corrections
(TMCs) \cite{Wandzura-tmc, BT-tmc, Matsuda80, Piccione98, AM08},
both of which involve additional integrations in $x$ space.
In practice this is achieved by precomputing tables of moments
which can be retrieved during the computation of inverse
Mellin transforms.
Within this approach the TMCs can be evaluated to all orders
in $M^2/Q^2$, where $M$ is the nucleon mass, instead of just
including several low-order terms in the expansion \cite{JAM13}.

Another improvement in our new theoretical framework is in the
treatment of higher twist contributions to the spin-dependent
$g_1$ and $g_2$ structure functions.  In Ref.~\cite{JAM13} the
twist-3 part of $g_2$ was parametrized in terms of a light-cone
quark model inspired function of $x$ with 3 parameters, while the
twist-4 part of $g_1$ was fitted using a spline approximation for
the $x$ dependence of the $1/Q^2$ term, with knots for the spline
at several different $x$ values.
Here we adopt for both the twist-3 and twist-4 contributions
to $g_1$ and $g_2$ the same generic functional form as for the
leading twist PDFs, including for the first time a separation
into individual $u$ and $d$ flavors (we assume the higher twist
contributions, which are more relevant at large $x$ values,
to be small in the strange quark sector).
In addition, we include TMCs for the twist-3 distributions
\cite{BT-tmc}, along with the standard mass corrections for
the twist-2 PDFs, as well as $Q^2$ evolution of the twist-3
functions \cite{Braun01, Braun09}.

As in the previous JAM analysis \cite{JAM13}, we use the measured
$A_\|$ and $A_\perp$ asymmetries, whenever available, instead of
the derived $A_1$ asymmetry or $g_1$ structure function to avoid
uncertainties associated with inconsistent use of spin-averaged
structure functions in the extraction of the spin-dependent
observables.
We include new data sets with high-precision $A_\|$ and $A_\perp$
asymmetry measurements at Jefferson Lab from the ``eg1b''
\cite{eg1b-p, eg1b-d} and \mbox{``eg1-dvcs''} \cite{eg1-dvcs}
analyses on the proton and deuteron, and new results from the
\mbox{E06-014} experiment on $^3$He from Hall~A
\cite{E06-014_A1, E06-014_d2}.
Also included are the most recent $A_1$ measurements on the
proton from COMPASS \cite{COMPASS16}.
To more directly isolate the impact of the new data sets and
assess the systematics of our new methodology, we restrict
the current analysis to inclusive DIS data only.
A full analysis of all data, including semi-inclusive DIS, and
inclusive jet and $\pi$ production in polarized $pp$ collisions,
will be presented in a forthcoming publication \cite{JAM16}.

In Sec.~\ref{s.theory} of this paper we present a brief review of
the basic observables in spin-dependent DIS, and summarize the
essential results for the $g_1$ and $g_2$ structure functions at
finite $Q^2$, including the effects of target mass, higher twist
and nuclear corrections.
Our fitting methodology is discussed in Sec.~\ref{s.methodology},
where we describe the Mellin space technique and the details of
the iterative Monte Carlo procedure.
Section~\ref{s.data} summarizes the data used in the current fit,
and the results of the global analysis are presented in
Sec.~\ref{s.results}.
Here we systematically study the stability of the results with
respect to cuts on the data for different minimum values of $W^2$
and $Q^2$, in order to establish the extent of the kinematics over
which the formalism can provide a reliable description of the data.
For the optimal cuts determined by the stability of the moments
and the $\chi^2$ values, we present in Sec.~\ref{ss.comparisons} a
detailed comparison of the fitted results with all of the measured
polarization asymmetries from the earlier and new experiments.

The impact of the new Jefferson Lab data on the PDFs and their
uncertainties is discussed in Sec.~\ref{ss.jlab}, including the
most precise determination to date of the $x$ dependence of the
twist-3 distributions.  
The extracted twist-2 and twist-3 JAM15 PDFs are presented in
Sec.~\ref{ss.PDFs}, along with the fitted residual higher twist
contributions to the structure functions, including the $Q^2$
dependence of the $d_2$ moments of the twist-3 distributions.
Finally, in Sec.~\ref{s.conclusion} we summarize our results and
preview future extensions of this work.

\section{Formalism}
\label{s.theory}

In this section we give a brief review of the basic framework for
polarized DIS, including the formulas for the measured polarization
asymmetries, and the essential results for the spin-dependent
structure functions in the operator product expansion of QCD.
We also review the unpolarized structure function input that is
needed for the extraction of the spin-dependent PDFs from the
measured asymmetries.

\subsection{Observables}
\label{ss.asymmetries}

The inclusive polarized DIS experiments used in this analysis measured
cross section asymmetries for lepton scattering from a stationary target
with various combinations of target and lepton spin, with the latter
always aligned or antialigned with the direction of the lepton beam.
While some experiments also measured absolute cross section differences
\cite{E06-014_A1, E06-014_d2, E01-012}, here we only use the
polarization asymmetries.

In the most general case, with the target polarization pointing in a
direction given by spherical polar angles $\theta^*$ and $\phi^*$
relative to the direction of the momentum transfer vector $\bm{q}$,
the measured asymmetry is defined as
\begin{align}
A\ &=\
\frac{\sigma^{\downarrow}-\sigma^{\uparrow}}
     {\sigma^{\downarrow}+\sigma^{\uparrow}}
\ =\ \frac{\cos\theta^* \sqrt{1-\epsilon^2} A_1
	  +\sin\theta^* \cos\phi^* \sqrt{2 \epsilon(1-\epsilon)} A_2}
	  {1+\epsilon R},
\label{e.Agen}
\end{align}
where the arrow $\uparrow$ ($\downarrow$) denotes the spin of the
lepton along (opposite to) the beam direction.  The variable
\begin{align}
\epsilon 
=\frac{2(1-y) - \frac{1}{2}\gamma^2 y^2}{1+(1-y)^2 + \frac{1}{2}\gamma^2 y^2}
\label{e.eps}
\end{align}
is the ratio of longitudinal to transverse photon polarizations,
where $y = \nu/E$ is the fractional energy transfer from the
lepton in the target rest frame, $\gamma^2 = 4 M^2 x^2/Q^2$,
and $x=Q^2/2M\nu$ is the Bjorken scaling variable.
In Eq.~(\ref{e.Agen}), $A_1$ and $A_2$ are the virtual photoproduction
asymmetries, and $R = \sigma_L/\sigma_T$ is the ratio of the
longitudinal to transverse virtual photoproduction cross sections.
For the case where the target polarization is either along
($\Uparrow$) or perpendicular to ($\Rightarrow$) the beam direction,
the general expression for the asymmetry in Eq.~(\ref{e.Agen})
reduces to the longitudinal and transverse asymmetries, defined by
\begin{align}
A_\|&=
\frac{\sigma^{\downarrow \Uparrow}-\sigma^{\uparrow \Uparrow}}
     {\sigma^{\downarrow \Uparrow}+\sigma^{\uparrow \Uparrow}}
= D (A_1 + \eta A_2),				\label{e.Apar}  \\
A_\perp&=
\frac{\sigma^{\downarrow \Rightarrow}-\sigma^{\uparrow \Rightarrow}}
     {\sigma^{\downarrow \Rightarrow}+\sigma^{\uparrow \Rightarrow}}
= d (A_2 - \zeta A_1),				\label{e.Aperp}
\end{align}
where the kinematical variables here are given by
\begin{align}
D &= \frac{y(2-y)(2+\gamma^2 y)}
	  {2(1+\gamma^2)y^2+(4(1-y)-\gamma^2y^2)(1+R)},	\nonumber\\
d &= \frac{\sqrt{4(1-y)-\gamma^2 y^2}}{2-y}D,
\label{e.kinfact}					\\
\eta & = \gamma \frac{4(1-y)-\gamma^2y^2}{(2-y)(2+\gamma^2 y)},\ \ \ \
\zeta = \gamma \frac{2-y}{2+\gamma^2y}.		\nonumber
\end{align}
These definitions for the asymmetries are consistent with the ones
commonly found in the literature (in which the spin of the lepton is
fixed but that of the target is flipped), if parity-violating
effects can be neglected.
%
%
The virtual photoproduction asymmetries can be expressed as ratios of
spin-dependent ($g_1$ and $g_2$) and spin-averaged ($F_1$ and $F_2$)
structure functions,
\begin{align}
A_1 &= \frac{(g_1-\gamma^2g_2)}{F_1},\ \ \ \ \ \
A_2  = \gamma\frac{(g_1+g_2)}{F_1},
\label{e.A}
\end{align}
with the ratio $R$ given in terms of the spin-averaged structure
functions by
\begin{align}
R &= \frac{ (1+\gamma^2)F_2 - 2xF_1 }{ 2x F_1 }.
\label{e.R}
\end{align}
At large values of $Q^2$, the variables $\eta$ and $\zeta$ in
Eq.~(\ref{e.kinfact}) vanish, and the longitudinal and transverse
asymmetries become proportional to $A_1$ and $A_2$, respectively.
In this case the polarization asymmetry $A_1 \approx g_1/F_1$,
and has a simple interpretation in terms of parton distributions,
as we discuss next.

\subsection{Structure functions in QCD}
\label{ss.sfs}

In the leading twist (twist $\tau=2$) approximation the $g_1$
structure function can be computed in terms of spin-dependent PDFs,
\begin{align}
g_1^{(\tau 2)}(x,Q^2)
&=\frac{1}{2}
  \sum_q e_q^2
  \left[ (\Delta C_q \otimes \Delta q^+)(x,Q^2)
       +\ (\Delta C_g \otimes \Delta g)(x,Q^2)
  \right],
\end{align}
where $\Delta q^+ = \Delta q + \Delta \bar q$ is the sum of the
quark and antiquark PDFs, $\Delta g$ is the gluon PDF, and
$\Delta C_q$ and $\Delta C_g$ are the respective hard scattering
coefficients, calculable in perturbative QCD.  In this analysis we
use the hard scattering coefficients computed to next-to-leading
order (NLO) accuracy, as is standard in all global spin PDF
analyses.  The symbol ``$\otimes$'' denotes the convolution integral,
$(\Delta C \otimes \Delta f)(x)
 = \int_x^1 (dz/z) \Delta C(z) \Delta f(x/z)$.
In the leading twist approximation, the $g_2$ structure function
is given in terms of the twist-2 component of $g_1$ via the
Wandzura-Wilczek relation \cite{Wandzura77},
\begin{align}
g_2^{(\tau 2)}(x,Q^2)
&= -g_1^{(\tau 2)}(x,Q^2)
 + \int_x^1\frac{dz}{z} g_1^{(\tau 2)}(z,Q^2).
\label{e.WW}
\end{align}
Defining the $N$-th moments of the $g_{1,2}$ structure functions as
\begin{align}
\mell{g}_{1,2}(N,Q^2)
&= \int_0^1 dx\, x^{N-1}\, g_{1,2}(x,Q^2),
\label{e.g12mom}
\end{align}
one finds that the lowest $(N=1)$ moment of $g_2^{(\tau 2)}$
satisfies the Burkhardt-Cottingham (BC) sum rule \cite{BC70},
\begin{align}
\mell{g}_2^{(\tau 2)}(1,Q^2) = 0.
\label{e.BC}
\end{align}

While these results are, strictly speaking, valid in the Bjorken
limit ($Q^2 \to \infty$, $x$ fixed), at finite values of $Q^2$,
power-suppressed [${\cal O}(1/Q^2)$] corrections to the structure
functions can make important contributions in some kinematic regions.
The simplest of these are the target mass corrections, which in
the operator product expansion are associated with matrix elements
of twist-2 operators with insertions of covariant derivatives
\cite{Nachtmann73}.
These do not alter the twist classification, but lead to
corrections to the structure functions that scale with the
Nachtmann variable $\xi$, where \cite{Nachtmann73, Greenberg71}
\begin{align}
\xi = \frac{2x}{1+\rho},\ \ \ \ \ \ \rho^2 = 1 + \gamma^2.
\end{align}
For the target mass corrected $g_1$ structure function, one has
\cite{Wandzura-tmc, BT-tmc}
\begin{align}
g_1^{(\tau 2+\rm TMC)}(x,Q^2)
&= \frac{x}{\xi \rho^3}\, g_1^{(\tau 2)}(\xi,Q^2)
 + \frac{(\rho^2-1)}{\rho^4}
   \int_\xi^1 \frac{dz}{z}
   \left[ \frac{(x+\xi)}{\xi}
	- \frac{(3-\rho^2)}{2 \rho} \ln\frac{z}{\xi}
   \right] g_1^{(\tau 2)}(z,Q^2),
\label{e.g1LTTMC}
\end{align}
while the $g_2$ target mass corrected structure function is
given by
\begin{align}
g_2^{(\tau 2+\rm TMC)}(x,Q^2)
&= -\frac{x}{\xi \rho^3}\, g_1^{(\tau 2)}(\xi,Q^2)
 + \frac{1}{\rho^4} 
   \int_\xi^1 \frac{dz}{z}
   \left[ \frac{x}{\xi}-(\rho^2-1)
	+ \frac{3(\rho^2-1)}{2\rho} \ln\frac{z}{\xi}
   \right] g_1^{(\tau 2)}(z,Q^2).
\label{e.g2LTTMC}
\end{align}
Note that in the presence of TMCs, the finite-$Q^2$ structure
functions in Eqs.~(\ref{e.g1LTTMC}) and (\ref{e.g2LTTMC})
are nonzero at $x=1$, vanishing only in the $\xi \to 1$ limit.
The nonvanishing of the target mass corrected structure functions
at $x=1$ is usually referred to as the ``threshold problem''
\cite{GP76, DGP77, DGP77annals}, and has been discussed at length
in the literature \cite{Bitar79, Steffens06, Schienbein08,
Steffens12, JMC}.
In practice, the kinematics where this problem becomes relevant
are restricted to the nucleon resonance region, at values of $W^2$
far below those where a perturbative QCD analysis is applicable.

The $Q^2$ dependence of the massless limit functions
$g_{1,2}^{(\tau 2)}$ on the right hand sides of Eqs.~(\ref{e.g1LTTMC})
and (\ref{e.g2LTTMC}) is due to the perturbative QCD evolution of the
twist-2 distributions themselves.
Clearly in the large-$Q^2$ limit, when $\rho \to 1$ and $\xi \to x$,
Eq.~(\ref{e.g2LTTMC}) reduces to the Wandzura-Wilczek relation,
Eq.~(\ref{e.WW}).
However, even in the presence of TMCs, Eq.~(\ref{e.WW}) with
$g_{1,2}^{(\tau 2)}$ replaced by $g_{1,2}^{(\tau 2+\rm TMC)}$
is still satisfied, provided the integration of the second term
is extended to $1/(1-M^2/Q^2)$, which corresponds to evaluating
the target mass corrected structure functions in
Eqs.~(\ref{e.g1LTTMC}) and (\ref{e.g2LTTMC}) up to $\xi=1$.
Moreover, the BC sum rule is also satisfied for the target mass
corrected structure function $g_2^{(\tau 2+\rm TMC)}$.

In addition to the kinematical TMCs, structure functions in the
operator product expansion receive contributions also from higher
twist terms which are associated with matrix elements of quark-gluon
or multi-quark operators.  As with the TMCs, these vanish at large
$Q^2$, but at low $Q^2$ values ($Q^2 \sim 1$~GeV$^2$) can play
an important role in DIS.  Of course, if $Q^2$ is too small,
then the expansion in $1/Q^2$ will not be convergent; however,
at low, but not too low, $Q^2$ values there will be a window in
which the higher twist contributions themselves can be extracted
from data \cite{Ji97, Deur04, Osipenko05, Meziani05}.
Keeping only the higher twist terms that contribute at the lowest
order in $\sim 1/Q^2$, we use the following expansion for the
structure functions,
\begin{align}
g_1 &= g_1^{(\tau 2)} + g_1^{(\tau 3)} + g_1^{(\tau 4)},
\label{e.g1ope} \\
g_2 &= g_2^{(\tau 2)} + g_2^{(\tau 3)},
\label{e.g2ope}
\end{align}
where, with the exception of the twist $\tau=4$ term, each of the
other ($\tau=2$ and 3) contributions implicitly contains TMCs.
In particular, for the twist-3 part of the $g_1$ structure function,
one has \cite{BT-tmc}
\begin{align}
g_1^{(\tau 3+\rm TMC)}(x,Q^2)
&= \frac{(\rho^2-1)}{\rho^3}\, D(\xi,Q^2)
 - \frac{(\rho^2-1)}{\rho^4}
   \int_\xi^1 \frac{dz}{z}
   \left[ 3 - \frac{(3-\rho^2)}{\rho} \ln\frac{z}{\xi}
   \right] D(z,Q^2),
\end{align}
where the function $D$ is expressed in terms of twist-3
parton distributions,
\begin{align}
D(x,Q^2) = \sum_q e_q^2 D_q(x,Q^2).
\label{e.D}
\end{align}
Similarly, for the target mass corrected twist-3 part of the $g_2$
structure function one has \cite{BT-tmc}
\begin{align}
g_2^{(\tau 3+\rm TMC)}(x,Q^2)
&= \frac{1}{\rho^3}\, D(\xi,Q^2)
 - \frac{1}{\rho^4}
   \int_\xi^1 \frac{dz}{z}
   \left[ 3 - 2 \rho^2 + \frac{3(\rho^2-1)}{\rho} \ln\frac{z}{\xi}
   \right] D(z,Q^2).
\label{e.g2t3}
\end{align}
Note that at large $Q^2$ the twist-3 part of $g_1$ vanishes,
since nonzero values of $g_1^{(\tau 3+\rm TMC)}$ arise only from
target mass effects.  On the other hand, the twist-3 part of the
$g_2$ structure function remains nonzero even in the $M^2/Q^2 \to 0$
limit (in which $\rho \to 1$ and $\xi \to x$), where it is given by
an expression similar to the Wandzura-Wilczek relation for the
twist-2 part of $g_2$,
\begin{align}
g_2^{(\tau 3)}(x,Q^2)
&= D(x,Q^2) - \int_x^1\frac{dz}{z}D(z,Q^2).
\label{e.g2t3_Bj}
\end{align}
In this limit, one can see by inspection that the \mbox{twist-3}
component of $g_2$ also satisfies the BC sum rule (\ref{e.BC}),
$\mell{g}_2^{(\tau 3)}(1,Q^2)=0$.
As in the case of the twist-2 contribution, the BC sum rule
also holds for the twist-3 part in the presence of TMCs.

In Eqs.~(\ref{e.g2t3}) and \eqref{e.g2t3_Bj} the $Q^2$ dependence
of the twist-3 function $D$ is generated perturbatively
\cite{Braun01, Braun09}, and in our analysis we use the
large-$N_c$ approximation to describe the evolution of the moments
$\mell{D}(N,Q^2)$ of the twist-3 functions in Mellin space,
\begin{align}
\mell{D}(N,Q^2)
\approx \left( \frac{\alpha_S(Q^2)}{\alpha_S(Q^2_0)}
	\right)^{\widetilde{\gamma}} \mell{D}(N,Q_0^2),
\label{e.Devol}
\end{align}
where the moments $\mell{D}(N,Q^2)$ are defined analogously to
Eq.~(\ref{e.g12mom}).
Here $\alpha_S$ is the strong running coupling, and the evolution
from the initial scale $Q_0^2$ is governed by the anomalous
dimension 
\begin{align}
\widetilde{\gamma}
= \frac{1}{(11-\frac{2}{3} N_f)}
  \left( \psi(0,N) + \gamma_E - \frac{1}{4} + \frac{1}{2N} \right),
\end{align}
where $\psi(0,N)$ is the polygamma function of order 0,
$\gamma_E$ is the Euler-Mascheroni constant, and $N_f$ is the
number of active flavors.

Of particular interest is the $d_2$ integral, which is defined by
a combination of $N=3$ moments of $g_1$ and $g_2$ \cite{Jaffe91},
\begin{align}
d_2(Q^2) = 2 \mell{g}_1(3,Q^2) + 3 \mell{g}_2(3,Q^2).
\label{e.d2}
\end{align}
From Eq.~(\ref{e.WW}) one observes that the twist-2 contributions
to $d_2$ vanish identically, so that the leading contributions to
$d_2$ arise at the twist-3 level.  
In terms of moments of the $D_q$ distributions in Eq.~(\ref{e.D}),
the leading (twist-3) part of $d_2$ is given by
\begin{align}     
d_2^{(\tau 3)}(Q^2) = \sum_q e_q^2\, \mell{D}_q(3,Q^2).
\label{e.d2tw3}
\end{align}    
Physically, $d_2$ is related to matrix elements describing the
nucleon's ``color polarizability'' \cite{Unrau94, Stein_t3, Stein_t4}
or the ``transverse color force'' \cite{Burkardt13} acting on quarks.

Finally, for the residual twist-4 and higher contributions to the
$g_1$ structure function in Eq.~(\ref{e.g1ope}) we use an effective
hadronic level parametrization, 
\begin{align}
g_1^{(\tau 4)}(x,Q^2) = \frac{H(x,Q^2)}{Q^2},
\label{e.g1t4}
\end{align}
where $H$ is in general a function of $x$ and $Q^2$.
Since the function $H$ will be fitted phenomenologically, and treated
as a background to the twist-2 and twist-3 contributions that are the
primary focus of our analysis, we do not include target mass or $Q^2$
evolution corrections in $H$.
For completeness, we also define the third moment of $H$ by
\begin{align}
h(Q^2) = \mell{H}(3,Q^2),
\label{e.h3}
\end{align}
where the Mellin transform $\mell{H}(N,Q^2)$ is defined as in
Eq.~(\ref{e.g12mom}).
In summary then, our analysis of the $g_1$ and $g_2$ structure
functions will involve the twist-2 polarized PDFs $\Delta q$ and
$\Delta g$, the twist-3 distributions $D_q$, and the residual
higher twist functions $H_{p,n}$ for the proton and neutron.

\newpage
\subsection{Spin-averaged structure functions}
\label{ss.F12}

The extraction of spin-dependent PDFs from the polarization asymmetries
in Sec.~\ref{ss.asymmetries} requires information on the spin-averaged
structure functions in the denominators of the asymmetries.
Ideally, the unpolarized and polarized structure functions should be
determined in a simultaneous fit to all DIS and other high energy
scattering data, to take into account the possible influence of the
spin-dependent data on the unpolarized observables.  Such correlations
are likely to be small, however, compared with the current uncertainties
on the asymmetries, and are neglected in all existing global PDF analyses.

In the JAM15 analysis we use the CJ12 global fit \cite{CJ12} of the
spin-averaged PDFs, taking advantage of the similarity in the DIS
kinematic cuts employed in both analyses, and the theoretical
treatment of target mass, higher twist and nuclear corrections.
The fitted CJ12 PDF parameters are then used to evolve the
unpolarized distributions and compute the spin-averaged structure
functions at the needed $Q^2$ scale.
In the CJ12 fit the strong coupling constant is computed using an
approximate analytical form, while the JAM15 analysis solves for
$\alpha_S$ numerically.  To avoid spurious numerical effects in the
calculation of the unpolarized structure functions from a mismatch
in the $Q^2$ evolution \cite{PEGASUS}, the CJ12 PDFs are refitted
utilizing the same numerical evolution routine adopted in the JAM
framework, and benchmarked against the natively calculated CJ12
observables.

The CJ12 analysis \cite{CJ12} provided NLO fits to the leading twist
PDFs, as well as the twist-4 contributions to the $F_2$ structure
function.  On the other hand, the polarization asymmetries in
Eq.~(\ref{e.A}) depend on the $F_1$ structure function, which
can be written as a combination of $F_2$ and the ratio $R$ in
Eq.~(\ref{e.R}).
Following Alekhin {\it et al.} \cite{Alekhin07}, who found very
small higher twist contributions to $R$ over the entire $x$ range
of the available DIS data, we set the twist-4 component of $R$ to
zero.  This allows the twist-4 part of $F_1$ to be computed as
  $F_1^{(\tau 4)} = F_1^{(\tau 2)}(1 + C_{\rm HT}(x)/Q^2)$,
with the higher twist $C_{\rm HT}(x)$ coefficient function taken
from the CJ12 fit for $F_2$ \cite{CJ12}.

For the TMCs, the CJ12 fit utilized the collinear factorization
formalism of Ref.~\cite{JMC} rather than the operator product
approach adopted here.  The differences, however, between the two
approaches have been shown \cite{Brady11} to be minimal in the
$x$ and $Q^2$ region covered by the spin-dependent data.

\section{Methodology} 
\label{s.methodology}

Having defined the polarization observables and structure functions
necessary for a QCD-based analysis, in this section we outline our
methodology for fitting the spin-dependent PDFs to the inclusive
DIS data.
We perform our analysis in moment space, which requires efficient
computation of inverse Mellin transforms, but has the advantage of
significantly shorter fitting times compared with $x$-space based
analyses \cite{DSSV09}.  Following this we describe the novel
aspect of our analysis, namely the iterative Monte Carlo technique.

\subsection{Mellin space techniques}
\label{ss.mellin-space}

Calculation of the asymmetries and structure functions discussed
in the previous section involves at least two integrations for
both twist-2 and twist-3 observables.
For instance, the computation of the target mass corrected
$g_1^{(\tau 2+\rm TMC)}$ structure function involves a
convolution of the spin-dependent PDFs with the hard coefficient
functions, as well as additional integrations from the TMCs.
The numerical complexity of the problem further increases as
one considers the $Q^2$ evolution equations for the twist-2
distributions.

It turns out, however, that the computational burden can be
significantly reduced through the use of Mellin space techniques
\cite{DSSV09}.  Firstly, the $Q^2$ evolution equations in Mellin
space are ordinary coupled differential equations, which are
simpler and faster to solve compared with the corresponding
integro-differential equations in $x$-space.
Secondly, using the techniques developed by Stratmann and
Vogelsang \cite{Stratmann01}, it is possible to cast the various
multidimensional integrations in terms of precomputed quantities,
thereby significantly decreasing the computational time needed
for the observables in the global fits.

To illustrate the technique, consider the case of
$g_1^{(\tau 2+\rm TMC)}$ in \eref{g1LTTMC}.
For this we write the leading twist part of $g_1$ in the
Mellin representation as
\begin{align}
g_1^{(\tau 2)}(x,Q^2)
= \frac{1}{2\pi i}
  \int dN\, x^{-N} \mell{g}_1^{(\tau 2)}(N,Q^2),
\end{align}
where the moments $\mell{g}_1^{(\tau 2)}(N,Q^2)$ are defined in
Eq.~(\ref{e.g12mom}), and inserting this into the target mass
corrected expression in \eref{g1LTTMC} gives
\begin{align}
g_1^{(\tau 2+\rm TMC)}(x,Q^2)
= & \frac{1}{2\pi i}
  \int dN\, \mell{g}_1^{(\tau 2)}(N,Q^2)	\notag\\
& \times
  \left\{
  \frac{x}{\xi^{N+1} \rho^3}
+ \frac{(\rho^2-1)}{\rho^4}
  \int_\xi^1 \frac{dz}{z^{N+1}}
  \left[ \frac{(x+\xi)}{\xi}
       - \frac{(3-\rho^2)}{2\rho} \ln\frac{z}{\xi}
  \right]
  \right\}.
\label{e.mellin-trickI}
\end{align}
To simplify the notation we define the quantity in the braces
by $\mathcal{M}(x,Q^2,N) \equiv \left\{ \cdots \right\}$
in Eq.~(\ref{e.mellin-trickI}), which is a function of
$x$, $Q^2$ and $N$.
Crucially, $\mathcal{M}(x,Q^2,N)$ is independent of the
parameters to be fitted, which are confined entirely in the
$\mell{g}_1^{(\tau 2)}$ moments.
Furthermore, the moments $\mell{g}_1^{(\tau 2)}$ are simple
products of the moments of the hard coefficients and the
spin-dependent PDFs, so that \eref{mellin-trickI} can be
recast in the form
\begin{eqnarray}
g_1^{(\tau 2+\rm TMC)}(x,Q^2)
&=& \frac{1}{2\pi i}
  \int dN\, \mathcal{M}(x,Q^2,N)		\notag\\
& & \times
  \frac{1}{2} \sum_q e_q^2
  \Big[ \Delta\mell{C}_q(N) \Delta\mell{q}^+(N,Q^2)
      + \Delta\mell{C}_g(N) \Delta\mell{g}(N,Q^2)
  \Big].
\label{e.mellin-trickII}
\end{eqnarray}
Here the integration over $N$ is performed numerically in the
standard way by using a contour in the complex plane parametrized
as $N = c + z\, e^{i\phi}$.
The contour crosses the real axis at $c$, which is set to the right
of the rightmost pole of the integrand, and $\phi$ is set equal to
$3\pi/4$ to guarantee convergence of the integral.
Using the symmetry of the integrand with respect to the real
axis one can then write \eref{mellin-trickII} as
\begin{eqnarray}
g_1^{(\tau 2+\rm TMC)}(x,Q^2)
&=& \frac{1}{\pi}
  \int_0^\infty dz\,
  \text{Im} \Big\{ e^{i\phi} \mathcal{M}(x,Q^2,N)	\notag\\
& &\times
  \frac{1}{2} \sum_q e_q^2
  \Big[ \Delta\mell{C}_q(N) \Delta\mell{q}^+(N,Q^2)
      + \Delta\mell{C}_g(N) \Delta\mell{g}(N,Q^2)
  \Big]
  \Big\}.
\label{e.mellin-trickIII}
\end{eqnarray}
Expressing the integration over $z$ in terms of a Gaussian
quadrature sum with Gaussian weights $w_i$ \cite{PEGASUS},
one can approximate
\begin{eqnarray}
g_1^{(\tau 2+\rm TMC)}(x,Q^2)
&\simeq& \frac{1}{\pi} \sum_i w_i\,
  \text{Im} \Big\{ e^{i\phi} \mathcal{M}(x,Q^2,N_i) 	\notag\\
& &\times
  \frac{1}{2} \sum_q e_q^2
  \Big[ \Delta\mell{C}_q(N_i) \Delta\mell{q}^+(N_i,Q^2)
      + \Delta\mell{C}_g(N_i) \Delta\mell{g}(N_i,Q^2)
  \Big]
  \Big\},
\label{e.mellin-trickIIIgauss}
\end{eqnarray}
where now all the unknown quantities to be fitted (namely,
$\Delta q^+$ and $\Delta g$) decouple from the multidimensional
integrations which are contained inside $\mathcal{M}(x,Q^2,N_i)$.
The latter can be computed prior to the fit such that the
observable becomes a simple finite sum over the complex moments
$N_i = c + z_i\, e^{i\phi}$.

Potentially similar complications arise with the implementation
of the nuclear smearing corrections, in which the nuclear
(deuteron and $^3$He) structure functions are expressed as
convolutions of the nuclear smearing functions and bound
nucleon structure functions \cite{KM08, KM09, Ethier13},
\begin{align}
g_i^A(x,Q^2)
= \sum_{\tau=p,n}
  \int_x^A \frac{dz}{z}
  f_{ij}^{\tau/A}(z,\rho)\, g_j^\tau\left(\frac{x}{z},Q^2\right),
\end{align}
where the smearing function $f_{ij}^{\tau/A}(z,\rho)$ represents
the spin-depenent light-cone momentum distribution of nucleon
$\tau=p$ or $n$ in the nucleus $A$, and $g_j^\tau$ is the nucleon
structure function ($i,j=1,2$).  In principle the bound nucleon
structure functions can also depend on the degree to which the
nucleons are off-shell, but in practice these effects are likely
to be smaller than the current experimental uncertainties on the
polarization data \cite{Ethier13}.
At large $Q^2$ the smearing functions $f_{ij}^{\tau/A}$ are
steeply peaked around $z=1$ and are independent of $Q^2$,
but acquire $Q^2$ (or rather $\rho$) dependence at finite
$Q^2$ values \cite{KM08, KM09}.
In moment space the nuclear structure functions can also be
expressed in the compact form
\begin{align}
g_i^A(x,Q^2)
= \sum_{\tau=p,n}
  \frac{1}{2\pi i} \int dN\,
  \mathcal{M}^{\tau/A}_{ij}(x,Q^2,N)\, \mell{g}_j^{\tau}(N,Q^2),
\label{e.WB2}
\end{align}
where the smeared nuclear kinematic factor is given by
\begin{align}
\mathcal{M}^{\tau/A}_{ij}(x,Q^2,N)
= \int_0^1 \frac{dz}{z}
  f_{ij}^{\tau/A}(z,\rho)\, \mathcal{M}\left(\frac{x}{z},Q^2,N\right),
\label{e.WB}
\end{align}
which now contains both nuclear and target mass corrections.
As for the TMC implementation in
Eqs.~(\ref{e.mellin-trickII})--(\ref{e.mellin-trickIIIgauss}),
the factors $\mathcal{M}^{\tau/A}_{ij}$ can be precomputed,
allowing a more efficient evaluation of the nuclear structure
functions during the fitting procedure.

\subsection{PDF parametrization and errors}
\label{ss.parametrization}

For the generic parametrization of the spin-dependent PDFs,
as well as the twist-3 distributions $D_q$ and the twist-4
functions $H_{p,n}$, we choose the standard functional form
\begin{align}
f(x,Q_0^2) = {\cal N}\, x^a (1-x)^b (1 + c\sqrt{x} + d\, x)
\label{e.parametrization}
\end{align}
at the input scale $Q_0^2$, in terms of the four shape parameters
$a$, $b$, $c$ and $d$, and the normalization ${\cal N}$.
In Mellin space the moments of $f$ are defined as in
Eq.~(\ref{e.g12mom}) and can be expressed analytically 
using the beta function $B$,
\begin{eqnarray}
\mell{f}(N,Q_0^2)
&=&{\cal N}
  \Big[ B(N+a,b+1) + c\, B(N+a,b+3/2) + d\, B(N+a,b+2)
  \Big].
\label{e.mell-parametrization}
\end{eqnarray}
Since the present analysis only considers inclusive DIS data,
we attempt to fit only the PDFs $\Delta u^+$, $\Delta d^+$,
$\Delta s^+$ and $\Delta g$, and the higher twist distributions
$D_u$, $D_d$, $H_p$ and $H_n$.  For the polarized sea quark
distributions we follow some previous PDF analyses \cite{BB10}
in assuming a flavor symmetric sea,
\begin{align}
\Delta \bar{s}(x,Q^2)
= \Delta \bar{u}(x,Q^2)
= \Delta \bar{d}(x,Q^2)
= \frac{1}{2}\Delta s^+(x,Q^2).
\label{e.SU3}
\end{align}
Additional constraints on the moments of the PDFs are provided
by the weak neutron and hyperon decay constants,
\begin{align}
& \Delta\mell{u}^+(1,Q^2)
- \Delta\mell{d}^+(1,Q^2)	
= g_A,
\label{e.gA}			\\
& \Delta\mell{u}^+(1,Q^2)
+ \Delta\mell{d}^+(1,Q^2)
-2\Delta\mell{s}^+(1,Q^2)	
= a_8,
\label{e.a8}
\end{align}
where the moments $\Delta\mell{q}^+(1,Q^2)$ are defined as in
Eq.~(\ref{e.mell-parametrization}), and the triplet and octet
axial vector charges are given by $g_A = 1.269(3)$ and
$a_8 = 0.586(31)$, respectively.
Note that the nonsinglet combinations in Eqs.~(\ref{e.gA})
and (\ref{e.a8}) are independent of $Q^2$, whereas the quark
singlet combination,
\begin{align}
\Delta\Sigma(Q^2)
= \sum_q \Delta\mell{q}^+(1,Q^2),
\label{e.Sigma}
\end{align}
as well as the gluon moment
  $\Delta G(Q^2) = \Delta\mell{g}(1,Q^2)$,
are scale dependent.

The fit parameters are determined by minimizing the $\chi^2$
function, which we define as
\begin{align}
\chi^2 &= \sum_e
\left[ \sum_i
  \left( \frac{{\cal D}_i^{(e)} N^{(e)}_i - T^{(e)}_i}
	      {\alpha^{(e)}_iN^{(e)}_i}
  \right)^2
+ \sum_k \left(r^{(e)}_k\right)^2
\right],
\label{e.chi2}
\end{align}
where ${\cal D}_i^{(e)}$ is the measured value of the observable
for the data point $i$ from the experimental data set $e$,
with $T^{(e)}_i$ the corresponding theoretical value;
$\alpha^{(e)}_i$ represents the uncorrelated statistical
and systematic uncertainties added in quadrature.
To account for correlated (point-to-point) systematic
uncertainties $\beta^{(e)}_{k,i}$ in each experiment $e$,
we introduce normalization factors of the form
\begin{align}
N^{(e)}_i
&= 1 - \frac{1}{{\cal D}_i^{(e)}} \sum_k r^{(e)}_k \beta^{(e)}_{k,i},
\label{e.norm}
\end{align}
parametrized in terms of ``nuisance parameters'' $r^{(e)}_k$.
To control the size of the normalization factors, a penalty term 
is introduced in Eq.~(\ref{e.chi2}) as a quadrature sum of the
$r_k^{(e)}$ values, such that the fitted normalization factors
resemble Gaussian statistics.

Note that the normalization factors $N^{(e)}_i$ rescale both the
data values ${\cal D}_i^{(e)}$ and the uncorrelated (statistical
and point-to-point systematic) uncertainties $\alpha^{(e)}_i$.
This accounts for the fact that overall experimental scale factors,
such as the beam and target polarizations and dilution factors,
multiply both the data values and (in particular) their statistical
errors (which dominate $\alpha^{(e)}_i$) in the same way.
Moreover, considering only the rescaling of ${\cal D}_i^{(e)}$
would lead to a strong downward bias, known as D'Agostini bias
\cite{D'Agostini94}.

\subsection{Iterative Monte Carlo fitting}
\label{ss.IMC}

In standard single-fit PDF analyses, one often finds that some of
the shape parameters in Eq.~(\ref{e.parametrization}) are not well
determined by data and need to be fixed by hand, even when data
sets beyond inclusive DIS are considered \cite{DSSV09}.  This can
introduce additional arbitrariness into the analysis, since some
of the parameters and distributions have strong correlations.
Also, since the $\chi^2$ function is highly nonlinear in the fit
parameters, any single fit can find itself trapped in one of many
local minima, which only a Monte Carlo sampling can reveal.

\begin{figure}[t]
\includegraphics[width=0.85\textwidth]{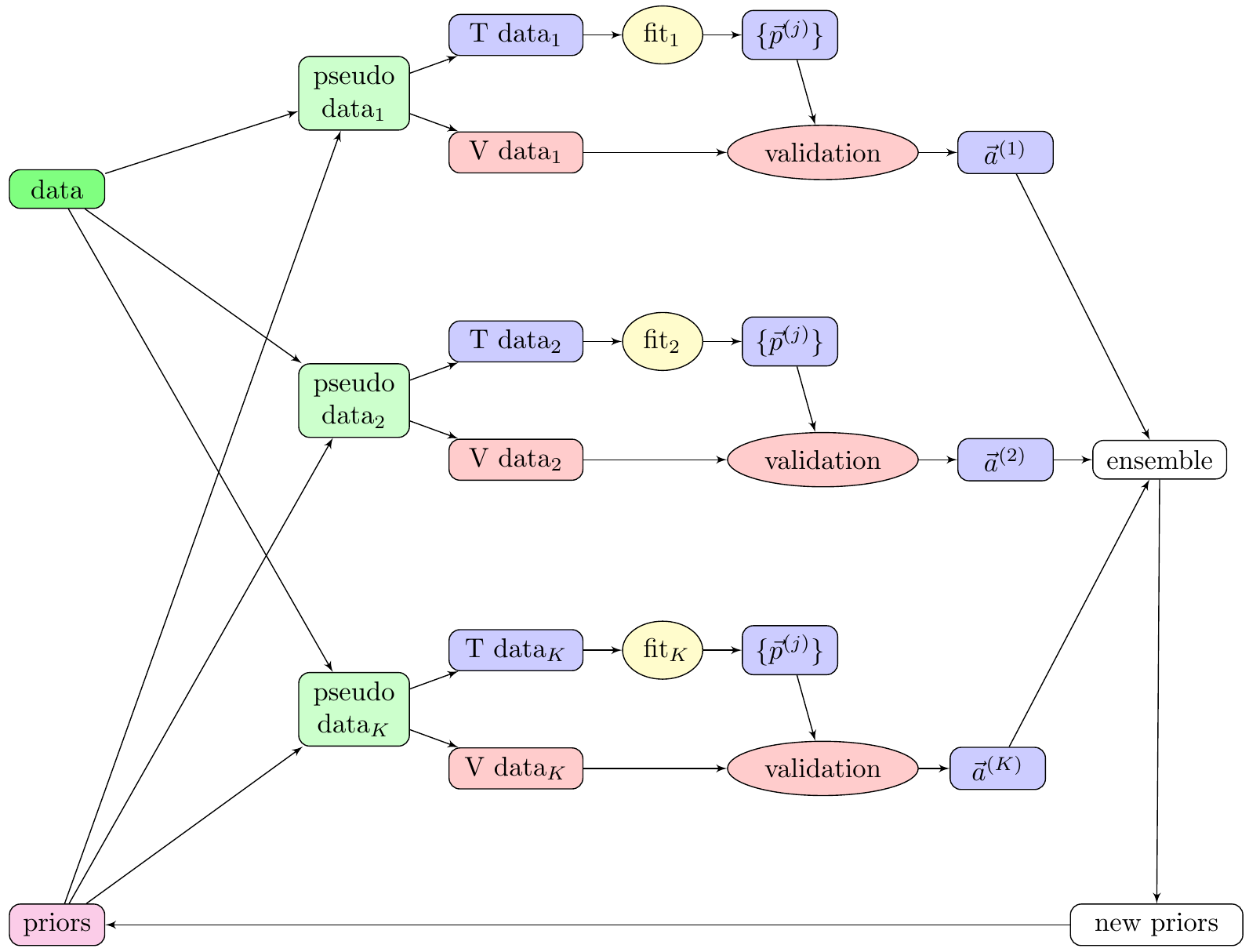}
\caption{Schematic illustration of the workflow for the iterative
	Monte Carlo fitting method.  In the first stage,
	$K$ pseudodata sets are generated, each of which is
	partitioned into training (T) and validation (V) subsets.
	For each pseudodata set, the training set is fitted and
	the parameters $\{ \vec{p}^{(j)} \}$ across all the
	minimization stages $j$ are stored.  The cross-validation
	procedure selects a single set of best fit parameters
	$\vec{a}^{(l)}$ from $\{ \vec{p}^{(j)} \}$ for each
	pseudodata set $l$, and the collection of
	$\{ \vec{a}^{(l)}; l=1,\ldots,K \}$ is then used as
	the priors for the next iteration.}
\label{f.workflow}
\end{figure}

For these reasons we have chosen instead to embark on a new approach
to global PDF analysis, based on an iterative Monte Carlo fitting
method that utilizes data resampling techniques and cross-validation.
Data resampling is used as a statistical error analysis method for
the extracted distributions as an alternative to the standard error
analysis using the Hessian method.
Cross-validation is a technique that prevents overfitting,
and is necessary in particular when using a large number
of fitting parameters.
The iterative procedure is summarized in \fref{workflow},
and involves the following key steps:
\begin{enumerate}
\item
{\bf Generation of pseudodata sets}: \\
Each pseudodata point is drawn from Gaussian sampling using the mean
and the uncertainties from the original experimental data values,
and is constructed as
\begin{align}
\widetilde{{\cal D}}_i = {\cal D}_i + R_i\, \alpha_i,
\end{align}
where ${\cal D}_i$ is an actual experimental data point,
$\alpha_i$ is the quadrature sum of the uncorrelated uncertainties,
and $R_i$ is a random number distributed from the normal distribution.
A total of $K$ pseudodata sets are generated this way.
\item
{\bf Partition of pseudodata sets}: \\
Each pseudodata set is partitioned randomly into ``training''
and ``validation'' sets using a splitting fraction of 50\%/50\%.
The partition of the data is performed within each experimental
data set to avoid experiments with few data points not appearing
in many of the fits.  Data sets with fewer than 10 points are not
partitioned, and are included as part of the training set.
\item
{\bf Generation of the priors}: \\
The priors are the set of parameters to be used as the starting
points for the fits.  During the initial iteration the priors for
each fit are generated using flat sampling of the parameter space
within a sufficiently broad region.  The ensemble of fitted
parameters or ``posteriors'' $\vec a^{(l)}$, with $l=1,\ldots,K$,
is then used as the priors for the next iteration.
\item 
{\bf $\chis$ minimization and cross-validation}: \\
The Levemberg-Marquardt gradient search algorithm \texttt{lmdiff}
\cite{More80} is used to minimize the $\chis$ function of the
training data set.  Information on the parameters $\{\vec{p}^{(j)}\}$
and the $\chis$ values of the training and validation sets across
each minimization stage $j$ is recorded.  The best fit parameters are
selected from the stage in which the lowest value in the validation
$\chis$ is attained.
\end{enumerate}
%

As mentioned earlier, the essential idea behind the iterative method is
to systematically transform the priors from the initial flat sampling
into posteriors that are distributed consistently with the information
contained within the data.  To assess the convergence of the posterior
distributions we examine the convergence of the corresponding
$\chisdof$ distribution.   In practice, the rate of convergence is
rather slow if one uses the full set of posteriors from one iteration
to the next.  To increase the efficiency of the iterative procedure,
in practice we select a subset of the posteriors that give the smallest
$\chisdof$ values, making a cut at the peak in the $\chisdof$ distribution
in a given iteration.  The signature of the convergence is then the
presence of an irreducible width in the $\chisdof$ distribution that
is generated from the selected sample of priors.

\begin{figure}[t]
\centering
\includegraphics[width=0.6\textwidth]{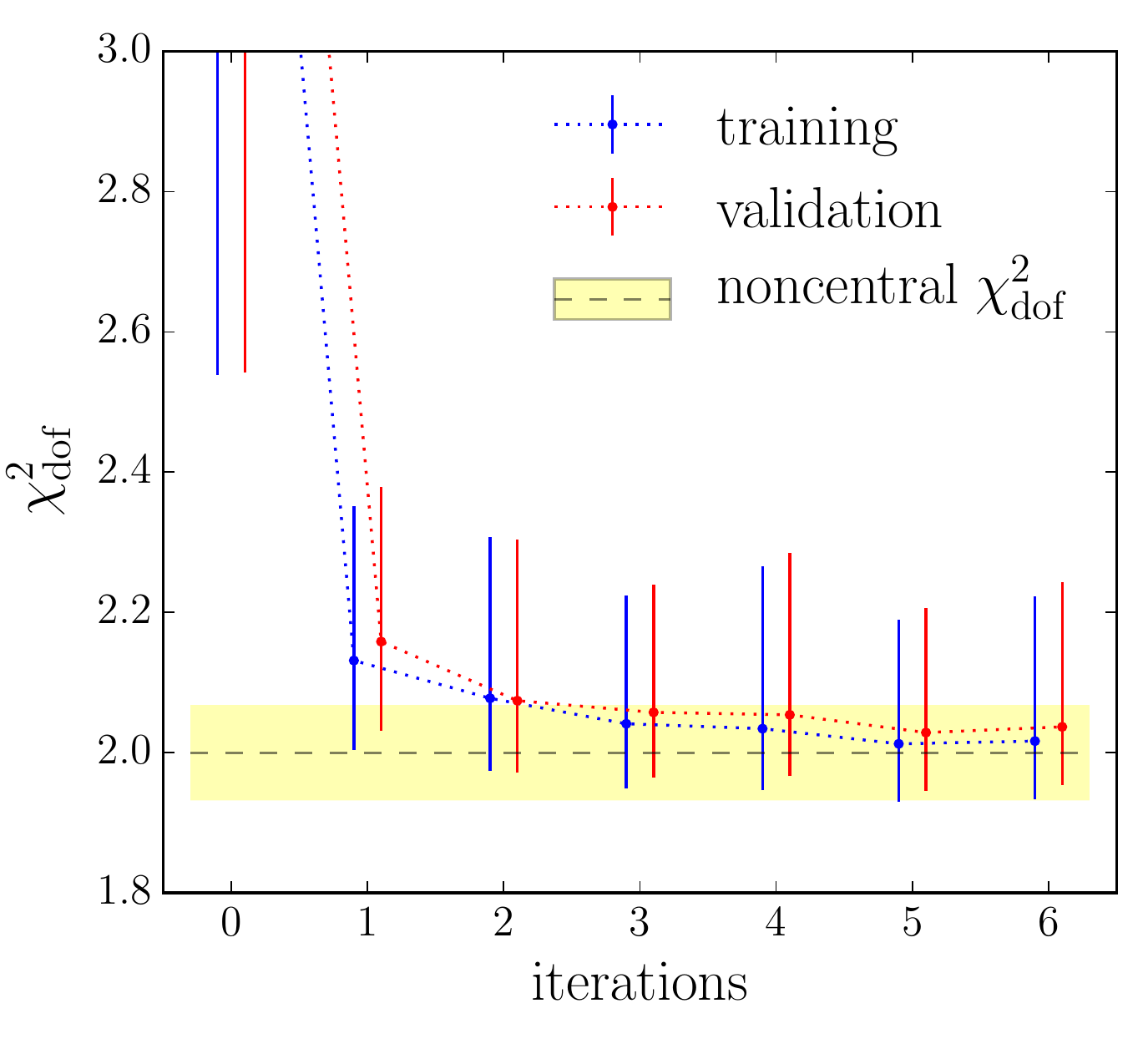}
\caption{Mean and two-sided standard deviations of the $\chisdof$
	distribution as a function of the iteration number for the
	training (blue points) and validation (red points) data sets,
	compared with the mean (dashed horizontal line at
	$\chisdof=2$) and standard deviation (yellow band)
	for the ideal noncentral $\chisdof$ distribution.}
\label{f.convergence-of-chi2} 
\end{figure}

In \fref{convergence-of-chi2} the mean and the two-sided standard
deviation of the training and validation $\chisdof$ distributions
are shown across the various iterations of the IMC procedure.
We find that statistical convergence of the $\chisdof$ distribution
is achieved after 5 or 6 iterations.
Notice that the $\chisdof$ distribution peaks around
$\chisdof \approx 2$, which is the expected behavior in the
idealized Gaussian statistics.  Namely, the $\chis$ values
obtained after fitting the many different realizations of the
data sets from the resampling are distributed according to the
noncentral $\chis$ distribution
\begin{align}
\mathcal{P}(\chis;n,\lambda)
= \frac{1}{2} \exp\left[-\frac{1}{2}(\chis+\lambda)\right]
  \left( \frac{\chis}{\lambda} \right)^{(n-2)/4}
  I_{n/2-1}(\sqrt{\lambda \chis}),
\label{e.noncentral-chi2}
\end{align}
where $I_{n/2-1}$ is the modified Bessel function of the first kind,
and $n$ is the number of degrees of freedom ($\approx$ number of data
points).  The parameter $\lambda$ is given by a sum of the expectation
values $\rm E$ of the individual point-by-point $\chi_i^2$ for the
data points,
  $\lambda = \sum_i^n {\rm E}[\chis_i]$.
In the ideal Gaussian statistics the expectation values are
  ${\rm E}[\chis_i] \simeq 1$,
and therefore $\lambda \simeq n$.
The noncentral $\chis$ distribution peaks around $2n$,
and the corresponding noncentral $\chisdof$ peaks around 2.

For comparison we also include in \fref{convergence-of-chi2} the
mean and standard deviation for the ideal noncentral $\chisdof$
distribution.  While the mean values of the IMC and ideal noncentral
$\chisdof$ distributions are in agreement, the right-side standard
deviation is generally larger for the IMC case.
This is somewhat consistent with the situation in the standard
error analysis in single fits, in which a tolerance in terms of
$\Delta\chis$ is defined in order to obtain conservative error
bands for the extracted PDFs.
We stress that in our approach the $\chisdof$ distribution is
extracted uniquely by the iterative procedure, and is determined
purely by the information contained in the data, thus removing the
need of any tolerance criterion.

The cross-validation in our procedure is implemented in two steps.
The first step is integrated within the iterative procedure and
corresponds to the selection of parameters from the minimization
steps, as described above.  The logic is that overfitting is
signaled whenever the $\chis$ of the training set continues to
improve across the minimization steps at the expense of a
deteriorating validation $\chis$.  The second step is implemented
once the statistical convergence of the posterior distribution is
attained.  We then examine each of the final posteriors
$\vec{a}^{(l)}$ by checking the difference in the $\chis$ values
between the validation and training sets.
A large difference also signals overfitting, which can occur if the
training set is not a statistically representative sample of the
entire data set, resulting in the partition creating an artificial
incompatibility within the data set itself.  The samples that are
ultimately selected are those that satisfy the condition
\begin{align}
\left| \chi^{2\, (\rm training)}_{\rm dof}
     - \chi^{2\, (\rm validation)}_{\rm dof}
\right| < 2\, \epsilon,
\end{align}
where $\epsilon$ is chosen to be the standard deviation of the ideal
noncentral $\chisdof$ distribution with the number of degrees of
freedom equal to the number of points in the training data set.

The final ensemble of posteriors is a collection of points in
the parameter space, each of which is represented by the vector
$\vec{a}^{(l)}$, whose components are the fitting parameters.
The distribution of the parameters is governed by the likelihood
function
\begin{align}
  \mathcal{P}(\vec{a}|{\cal D}) \propto 
  \exp\left[-\frac{1}{2}\chis(\vec{a})\right],
\end{align}
where $\chis$ is defined as in \eref{chi2}, and ${\cal D}$ denotes
the experimental data.  The ensemble of posteriors is therefore an
approximate Monte Carlo representation of the likelihood function
$\mathcal{P}(\vec{a}|{\cal D})$ for the fitting parameters $\vec{a}$.
The expectation values for the observables, such as a PDF at a
given $x$ and $Q^2$, can then be computed as 
\begin{align}
\rm E[\mathcal{O}]
&= \int d\vec{a}~\mathcal{P}(\vec{a}|{\cal D})~\mathcal{O}(\vec{a})
 = \frac{1}{K} \sum_l \mathcal{O}(\vec{a}^{(l)}).
\label{e.expectation}
\end{align}
In the last equality a Monte Carlo integration is performed by
sampling the parameters according to $\mathcal{P}(\vec{a}|{\cal D})$,
utilizing precisely the samples $\{\vec{a}^{(l)}; l=1,\ldots,K \}$
obtained after the IMC procedure.
Similarly, the variance of the observable can be computed as
\begin{align}
\rm V[\mathcal{O}]
&= \frac{1}{K} \sum_l
\left( \mathcal{O}(\vec{a}^{(l)}) -\rm E[\mathcal{O}] \right)^2,
\label{e.variance}
\end{align}
which gives the $1\sigma$ confidence interval for the observable
${\cal O}$.

Finally, in order to assess the goodness-of-fit, we also compute
the standard Pearson's $\chis$, defined as
\begin{align}
\chi^2
&=\sum_e
  \left[
    \sum_i
    \left( \frac{{\cal D}_i^{(e)} - {\rm E}[T^{(e)}_i/N^{(e)}_i]}
	 	{\alpha^{(e)}_i}
    \right)^2
  \right],
\end{align}
which differs slightly from the definition given in \eref{chi2}.
In particular, the actual data points ${\cal D}_i^{(e)}$ are used here
instead of the pseudodata points, and the theory values are computed
as expectation values in \eref{expectation}.  This definition allows a
direct comparison with $\chis$ values from single-fit based analyses.

\section{Data sets}
\label{s.data}

\begin{figure}[t]
\centering 
\includegraphics[width=0.7\textwidth]{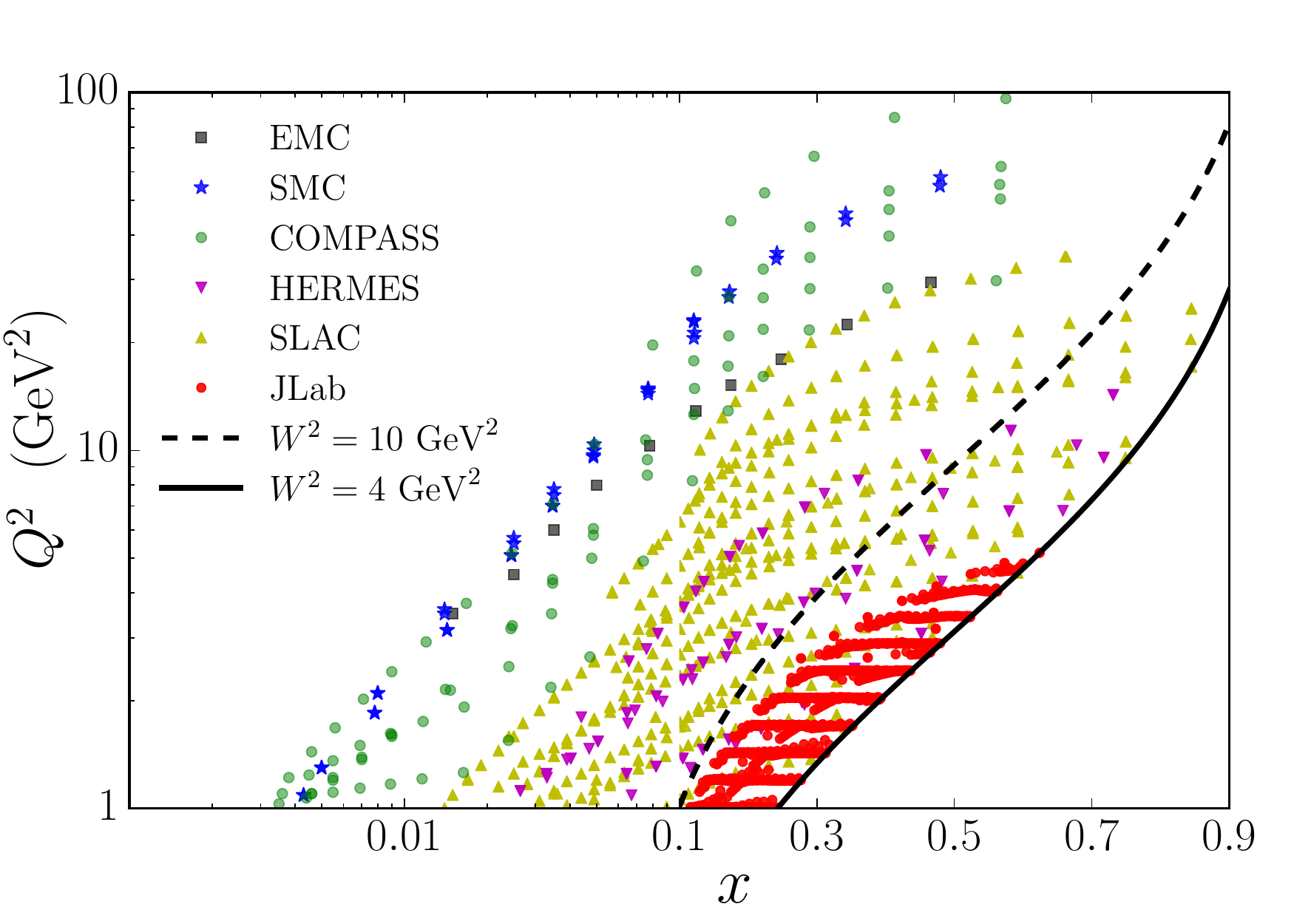}
\caption{Kinematic coverage in $x$ and $Q^2$ of the polarized
	inclusive DIS data sets used in the JAM15 analysis.
	The boundaries corresponding to fixed
	$W^2 = M^2 + Q^2 (1-x)/x$ equal to
	4~GeV$^2$ (solid curve)	and
	10~GeV$^2$ (dashed curve) are indicated.}
\label{f.kinematics} 
\end{figure}

The JAM15 global PDF analysis uses all available world data on
inclusive DIS of leptons (electrons, positrons and muons) on proton,
deuteron and $^3$He targets that pass the required cuts on the
invariant final state mass, $W^2 \geq 4$~GeV$^2$, and 
$Q^2 \geq 1$~GeV$^2$ (see Sec.~\ref{ss.cuts}).
This includes all of the sets from the
EMC \cite{EMC89},
SMC \cite{SMC98, SMC99}, COMPASS \cite{COMPASS10, COMPASS07},
SLAC \cite{SLAC-E130, SLAC-E142, SLAC-E143, SLAC-E154,
           SLAC-E155p, SLAC-E155d, SLAC-E155_A2pd, SLAC-E155x},
HERMES \cite{HERMES97, HERMES07, HERMES12}, and
Jefferson Lab Hall~A \cite{E99-117} experiments used in the
previous JAM13 global fit \cite{JAM13}, as well as the more recent
high-precision asymmetry measurements from Jefferson Lab
\cite{eg1b-p, eg1-dvcs, eg1b-d, E06-014_A1, E06-014_d2}
and new results from COMPASS \cite{COMPASS16}.
The data sets are summarized in Table~\ref{t.chi2}, and the kinematic
coverage in $x$ and $Q^2$ is illustrated in Fig.~\ref{f.kinematics}.
The Jefferson Lab data points are concentrated at intermediate
values of $x$ and $Q^2 \lesssim 5$~GeV$^2$, and are entirely excluded
by a $W^2 \geq 10$~GeV$^2$ cut, as is typically used in other PDF fits.
With the inclusion of the new Jefferson Lab results, the number of
data points more than doubles, from $\approx 1000$, considered in
the JAM13 fit, to $> 2500$ in the current analysis.

\begin{table}[t]
\caption{Inclusive DIS data sets used in the JAM15 global PDF analysis,
	indicating the observables fitted, the targets used, the
	number of data points in each experiment, and the respective
	$\chisdof$ values. \\}
\input{tab-chi2}

\label{t.chi2} 
\end{table}

A summary describing most of the earlier experiments from SLAC,
CERN, DESY and Jefferson Lab can be found in Ref.~\cite{Kuhn09};
here we give a few experimental details about the most recent
experiments from Jefferson Lab \cite{eg1b-p, eg1b-d, eg1-dvcs,
E06-014_A1, E06-014_d2} and COMPASS \cite{COMPASS16}.
All of these experiments can be considered continuations of
the extensive experimental programs of the Hall~A and CLAS
collaborations at Jefferson Lab and COMPASS at CERN.

\begin{itemize}

\item{\bf eg1b} 
Experiment eg1b was the second installment of the eg1 run group in
Jefferson Lab's Hall B and ran in 2000--2001.  It used the CLAS
spectrometer and proton ($^{15}$NH$_3$) and deuteron ($^{15}$ND$_3$)
targets polarized along the direction of the incoming electron beam to
measure the double spin asymmetry $A_\|$ in \eref{Apar}.  A
first round of publications \cite{eg1a,eg1b-p-Prok} from this
experiment focused on the results from the lowest (1.6~GeV) and
highest (5.8~GeV) beam energies.  In the meantime, the complete data
set (including data with 2.5 and 4.2~GeV beam energy) has been
analyzed, including numerous improvements in the procedures used to
correct for backgrounds, beam and target polarization, electromagnetic
radiative corrections, and kinematic reconstruction.  The final
results from eg1b for the deuteron have been published \cite{eg1b-d}
and the results for the proton (used in the present analysis)
will be published shortly \cite{eg1b-p}.  Due to the wide range
in beam energies and running conditions, eg1b covers the largest
range in $x$ and $Q^2$ of any experiment at Jefferson Lab.

\item{\bf eg1-dvcs}
As the last spin structure function measurement with CLAS in the
6~GeV era of Jefferson Lab, experiment eg1-dvcs ran in 2009 with
a significantly improved polarized target ($^{14}$NH$_3$ and
$^{14}$ND$_3$ polarized along the beam direction) at the highest
beam energy ($5.8-6$~GeV) available at the time.
This experiment differs from eg1b chiefly due to its much higher
integrated luminosity and a significantly larger minimum scattering
angle, yielding a much higher statistical precision in the DIS region.
Its results have been published in Ref.~\cite{eg1-dvcs}.
 
\item{\bf E06-014}
Experiment E06-014 ran in Hall~A of Jefferson Lab in 2009 with
the primary purpose of determining the higher twist moment
$d_2(Q^2)$ in Eq.~(\ref{e.d2}) for the neutron.  It measured
both parallel and transverse double spin asymmetries as in
Eqs.~(\ref{e.Apar}) and (\ref{e.Aperp}), as well as cross section
differences for electron scattering off $^3$He targets polarized
up to 50\% through spin-exchange optical pumping.  The use of two
beam energies (4.7 and 5.9~GeV) and the ``BigBite'' large acceptance
spectrometer resulted in a broad coverage of the DIS region for both
$d_2$ \cite{E06-014_d2} and $A_1$ \cite{E06-014_A1}.

\item{\bf COMPASS}
The final results of the 2011 run of the COMPASS experiment with
a 200~GeV muon beam and a longitudinally polarized proton (NH$_3$)
target have recently been published \cite{COMPASS16}.
Only the virtual photon asymmetry $A_1$ is given, but at the high
$Q^2$ of these data, corrections due to $A_2$ should be minimal.
COMPASS data provide the lowest accessible values for $x$ and the
largest $Q^2$ values for any given $x$, and are therefore very
important for the extraction of sea quark and gluon polarization
information from inclusive DIS data.

\end{itemize}

For all experiments where they are available, we fit directly the
measured asymmetries $A_\|$ [Eq.~(\ref{e.Apar})] and $A_\perp$
[Eq.~(\ref{e.Aperp})] rather than derived quantities, such as
$A_1$ and $A_2$.  The SLAC experiment E155x \cite{SLAC-E155x}
presents a special case, in that the target was not polarized
exactly at 90$^\circ$ relative to the beam direction, but
at 92.4$^\circ$.
In addition, the asymmetries were measured simultaneously by
three spectrometers, one of which was on the opposite side of
the beam line than the other two, which affects the definitions
of the angles $\theta^*$ and $\phi^*$ in Eq.~(\ref{e.Agen}).
Consequently, the average values of $\theta^*$ and $\phi^*$ must
be calculated for each kinematic bin, and Eq.~(\ref{e.Agen})
used to relate the measurement to the underlying physics
quantities in the fit.  The transverse asymmetry measured
in this experiment is therefore indicated by the symbol
$\widetilde{A}_\perp$ in Table~\ref{t.chi2} to differentiate
it from the usual $A_\perp$.

By far the largest number of data points (albeit in a limited
kinematic range --- see Fig.~\ref{f.kinematics}) is provided by
the eg1b \cite{eg1b-p, eg1b-d} and eg1-dvcs \cite{eg1-dvcs}
experiments, which account for nearly half of the total.
Due to the high statistical precision of these experiments
(especially eg1-dvcs), it is important to treat systematic
uncertainties properly in order to avoid unwarranted biases
in the fit.  As outlined in Sec.~\ref{ss.IMC}, we distinguish
between uncorrelated systematic uncertainties, which randomly
vary from one kinematic bin to the next, and correlated
systematic uncertainties, which change the normalization of
all data points from a given experiment by essentially the
same factor.  The former are added in quadrature to the
statistical uncertainties (yielding the total point-to-point
uncertainties $\alpha^{(e)}_i$ in Eq.~(\ref{e.chi2})), while
the latter are incorporated in the normalization factor
$N^{(e)}_i$ as defined in Eq.~(\ref{e.norm}).

For most experiments, the correlated systematic uncertainty is just
the  uncertainty on an overall normalization constant incorporating
the dilution factor and the beam and target polarization; in that
case the ratio $\beta^{(e)}_{k,i} / {\cal D}_i^{(e)}$ in
Eq.~(\ref{e.norm}) is simply a constant percentage which we take
from the quoted normalization uncertainty.
For the proton and deuteron data from the most recent CLAS
experiments \cite{eg1b-p, eg1b-d, eg1-dvcs}, a somewhat more
elaborate procedure is used, since an overall normalization
factor uncertainty is not available for these data.
In the case of eg1-dvcs \cite{eg1-dvcs}, the quoted systematic
uncertainties for all kinematic bins is completely dominated by
correlated normalization uncertainties.  Those quoted uncertainties
are therefore used directly for the quantity $\beta^{(e)}_{k,i}$
in Eq.~(\ref{e.norm}) (with the proper sign equal to that of the
data point in question and, since only one source of correlated
systematic error is quoted, $k=1$), without adding anything to the
statistical uncertainties.

For the proton data from eg1b \cite{eg1b-p}, only a small amount
of correlation, of order 3\% of the magnitude of the measured
asymmetry, is found between the systematic uncertainties for
different kinematic bins.  We therefore assign
  $\beta^{(e)}_{k=1,i} / {\cal D}_i^{(e)} = 0.03$
for all bins, but add the full systematic uncertainty in
quadrature to the statistical errors for $\alpha^{(e)}_i$. 
Finally, for the eg1b deuteron data set \cite{eg1b-d} one finds
a correlated systematic uncertainty of about 14\% for the 5.7~GeV
data ($\beta^{(e)}_{1,i} / {\cal D}_i^{(e)} = 0.14$) and 7\% for
the 4.2~GeV data ($\beta^{(e)}_{1,i} / {\cal D}_i^{(e)} = 0.07$).
Since this correlated part of the overall uncertainty is quite
sizable, it is subtracted from the quoted systematic uncertainties
in each bin.  The uncorrelated uncertainty
  $\sigma_{\rm uncor}
 = \sqrt{\sigma_{\rm tot\,\, sys}^2
        - \Large(\beta^{(e)}_{1,i}\Large)^2}$
is then added in quadrature to the statistical uncertainties.
In all cases the factors $r^{(e)}_k$ are optimized in the fit,
and the results indicate by which fraction of the correlated
uncertainties the data points of a given experiment have to be
moved to best agree with the world data.

\section{Results}
\label{s.results}

In this section we present the main results of the JAM15 global
analysis for the spin-dependent twist-2 and twist-3 distributions
and moments, and assess in particular the impact of the new
Jefferson Lab data on the PDFs and their uncertainties.
Before presenting the main results of the fits, we first examine
the dependence of the results on the kinematic cuts applied to the
data in order to maximize the range of $W^2$ and $Q^2$ over which
the data can be accommodated within our theoretical framework.

As mentioned above, for the initial iteration the priors for each
fit are generated from flat sampling of a reasonable range in the
parameter space.  While any restriction of the initial parameter
sampling in principle introduces a bias into the procedure,
we choose the parameter ranges to be sufficiently broad so as
to minimize any such bias, at the same time ensuring that the
parameters do not introduce unphysical behavior in any of the
observables.

Specifically, for the exponent $a$ governing the $x \to 0$ behavior of
the leading twist PDFs in Eq.~(\ref{e.parametrization}), we consider
the range $a \in [-1,0]$, which covers the values expected from Regge
theory, as well as the findings in all previous phenomenological PDF
analyses.  For the exponent $b$ that determines the $x \to 1$
behavior, we choose the range $b \in [2,5]$ for the $\Delta u^+$
and $\Delta d^+$ PDFs that have valence components at large $x$,
and $b \in [2,10]$ for the sea distributions $\Delta s^+$ and
$\Delta g$ that are more strongly suppressed as $x \to 1$.
In addition, we introduce penalties in the $\chis$ whenever the
$b$ parameter for $\Delta s^+$ or $\Delta g$ becomes lower than
the corresponding parameter for $\Delta d^+$.
For the auxiliary $c$ and $d$ shape parameters in
Eq.~(\ref{e.parametrization}), we set the starting ranges for both
between $-1$ and 1.  For the normalization of the singlet quark
and gluon first moments, we take the starting values such that
$\Delta \Sigma$ and $\Delta G$ are both equal to 0.5.

Considerably less is known about the shapes of the higher twist
distributions.  Generally, these are expected to play a greater
role at smaller $W$ values, or, for fixed $Q^2$, at large $x$.
To allow for additional suppression of the higher twists at small $x$,
we consequently take the initial range for the $a$ parameter for
the twist-3 and twist-4 functions to be $a \in [-1,1]$, with
normalization for all higher twists starting at zero.
For the large-$x$ parameter $b$ we take the initial sampling
region to be $b \in [2,5]$, and for the auxiliary parameters
$c, d \in [-1,1]$ for all higher twist distributions.

\subsection{$W^2$ and $Q^2$ cuts}
\label{ss.cuts}

\begin{table}
\caption{Dependence of the global fits on the cut on the hadronic
	final state mass squared, $W^2_{\rm cut}$, for a fixed
	$Q^2_{\rm cut} = 1$~GeV$^2$.  The $\chisdof$ values and
	the number of points included by the different $W^2$ cuts
	are listed, with the values for the JAM15 fit indicated
	in boldface. \\}
\begin{tabular*}{0.8\columnwidth}{@{\extracolsep{\fill}} lcccccc} \hline\hline
$W^2_{\rm cut}~(\rm GeV^2)$
		& 3.5  & {\bf 4}    & 5    & 6    & 8    & 10	\\ \hline
\# points       & 2868 & {\bf 2515} & 1880 & 1427 & 943  & 854	\\
$\chisdof$      & 1.20 & {\bf 1.07} & 1.03 & 1.02 & 0.99 & 0.97	\\ \hline
\label{t.tab-cuts-W}
\end{tabular*}
\end{table}

\begin{table}
\caption{Dependence of the global fits on the cut on the four-momentum
	transfer squared, $Q^2_{\rm cut}$, for a fixed
	$W^2_{\rm cut} = 4$~GeV$^2$.  The $\chisdof$ values and
	number of points included by the different $Q^2$ cuts are
	listed, with the JAM15 fit values indicated in boldface. \\}
\begin{tabular*}{0.5\columnwidth}{@{\extracolsep{\fill}} lccc} \hline\hline
$Q^2_{\rm cut}~(\rm GeV^2)$
		& {\bf 1.0}  & 2.0   & 4.0   	\\ \hline
\# points       & {\bf 2515} & 1421  & 611    	\\
$\chisdof$      & {\bf 1.07} & 1.08  & 0.95	\\ \hline
\label{t.tab-cuts-Q}
\end{tabular*}
\end{table}

To determine how far the kinematic boundaries delimited by the
$W^2$ and $Q^2$ cuts can be extended, we perform a series of
IMC fits, varying $W^2_{\rm cut}$ between 3.5 and 10~GeV$^2$
and $Q^2_{\rm cut}$ between 1 and 4~GeV$^2$.
The results of the fits are summarized in Tables~\ref{t.tab-cuts-W}
and \ref{t.tab-cuts-Q}, where the $\chisdof$ values are given,
along with the number of points included with each combination
of cuts.
For a fixed $Q^2_{\rm cut} = 1$~GeV$^2$, the number of points
more than triples when going from $W^2_{\rm cut}=10$~GeV$^2$ to
3.5~GeV$^2$, mostly due to the inclusion of the Jefferson Lab data,
but also because of important contributions from SLAC data.
Clearly, for the larger $W^2_{\rm cut}$ values very good fits can be
obtained with $\chisdof \approx 1$, which increases very gradually
as more data allowed by lower $W^2$ cuts are included in the fits.
For the lowest $W^2$ cut of 3.5~GeV$^2$, there is a somewhat larger
increase in the $\chisdof$ value.

\begin{figure}[t]
\centering 
\includegraphics[width=0.7\textwidth]{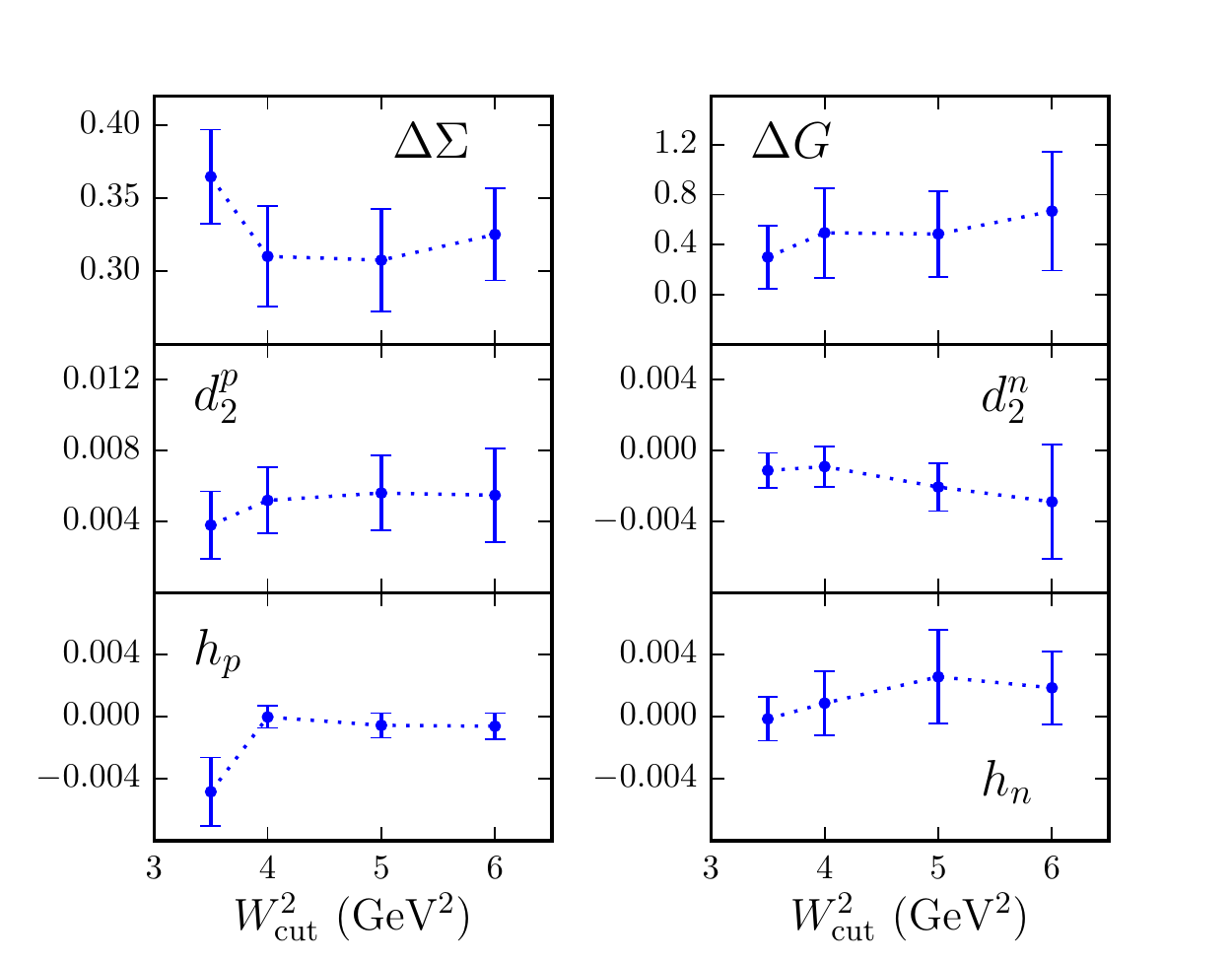}
\caption{Dependence on $W^2_{\rm cut}$ of several moments
	of twist-2 PDFs ($\Delta\Sigma$ and $\Delta G$),
	the twist-3 $d_2$ moments, and the third moments
	$h_p$ and $h_n$ of the twist-4 distributions of
	the proton and neutron.
	All fits use $Q^2_{\rm cut} = 1$~GeV$^2$, and the
	moments are truncated moments evaluated in the
	measured region between $x=0.001$ and 0.8.}
\label{f.mom_vs_W2}
\end{figure}

On the other hand, it is known that $\chis$ alone is not always
a sufficient indicator of the quality of the fit.
To examine the cut dependence in more detail, in
Fig.~\ref{f.mom_vs_W2} we show several moments of PDFs for
$W^2_{\rm cut}$ between 3.5 and 6~GeV$^2$, where the greatest
variations are expected to occur.  For higher values of
$W^2_{\rm cut}$, the results between 6 and 10~GeV$^2$ do not
change appreciably.  To avoid extrapolations into unmeasured
regions of $x$, we compute here the truncated moments,
evaluated between $x=0.001$ and 0.8, in the region covered
by the inclusive DIS data sets.  The lowest moment of the
twist-2 quark singlet distribution $\Delta\Sigma$ is found
to be rather stable down to $W^2_{\rm cut} = 4$~GeV$^2$,
increasing by $\sim 1\sigma$ at $W^2_{\rm cut} = 3.5$~GeV$^2$.
Similarly, the lowest moment of the gluon distribution $\Delta G$
is relatively flat as a function of $W^2_{\rm cut}$.

For the twist-3 $d_2$ proton and neutron moments, the variation
across $W^2_{\rm cut}$ is also fairly weak, although a significant
reduction in the uncertainty on the neutron $d_2^n$ is observed
when more of the low-$W^2$ data are included.  The impact of the
low-$W^2$ data is even more dramatically illustrated for the case
of the third moment of the twist-4 distribution of the proton
$H_p$, which shows a clear change in its central value between
$W^2_{\rm cut}=3.5$ and 4~GeV$^2$, and a significantly larger
uncertainty at the lower cut.  A stronger impact of low-$W^2$
data on higher twist contributions is not surprising, given that
higher twists are expected to be more important at larger $x$ values,
and the more rapid variation may be a signal of the presence of yet
higher twist corrections from the nucleon resonance region beyond
those considered in our analysis (see Sec.~\ref{ss.sfs}).

\begin{figure}[t]
\centering 
\includegraphics[width=0.7\textwidth]{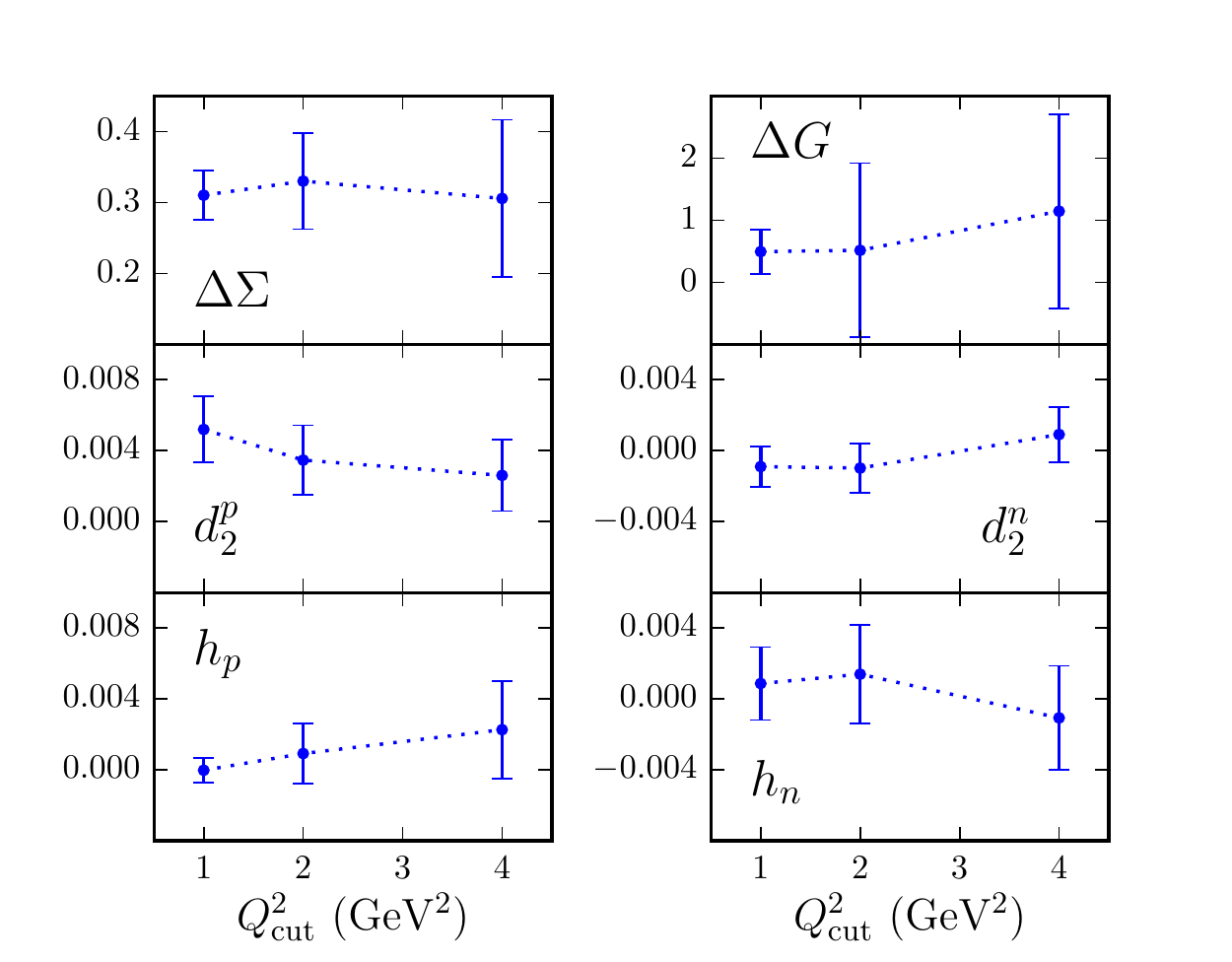}
\caption{As in Fig.~\ref{f.mom_vs_W2}, but for varying values
	of $Q^2_{\rm cut}$ between 1 and 4~GeV$^2$, for a
	fixed $W^2_{\rm cut} = 4$~GeV$^2$.}
\label{f.mom_vs_Q2}
\end{figure}

The dependence of the moments on the $Q^2$ cut is shown in
Fig.~\ref{f.mom_vs_Q2}.  The variation between $Q^2_{\rm cut}=1$
and 4~GeV$^2$ is generally mild and consistent within the errors.
Because of the reduced statistics for increasing values of
$Q^2_{\rm cut}$ (from $\sim 2500$ data points at 1~GeV$^2$ to
$\sim 1400$ points at 2~GeV$^2$, and $\sim 600$ at 4~GeV$^2$),
the uncertainties on the moments are correspondingly larger.
For the leading twist $\Delta\Sigma$ and $\Delta G$ moments,
for example, the uncertainties increase 3--4 fold between
$Q^2_{\rm cut}=1$ and 4~GeV$^2$.
With the aim of utilizing the maximum number of data points        
possible across all $W^2$ and $Q^2$ regions, while maintaining      
stable fits with good $\chi^2$ values, we therefore select
$W^2_{\rm cut} = 4$~GeV$^2$ and $Q^2_{\rm cut} = 1$~GeV$^2$
for the cuts to be used in the final JAM15 analysis.
All the results in the following sections will be based on
these values.

\subsection{Comparisons with experimental asymmetries}
\label{ss.comparisons}

\begin{figure}[bt]	
\centering 
\includegraphics[width=1.05\textwidth]{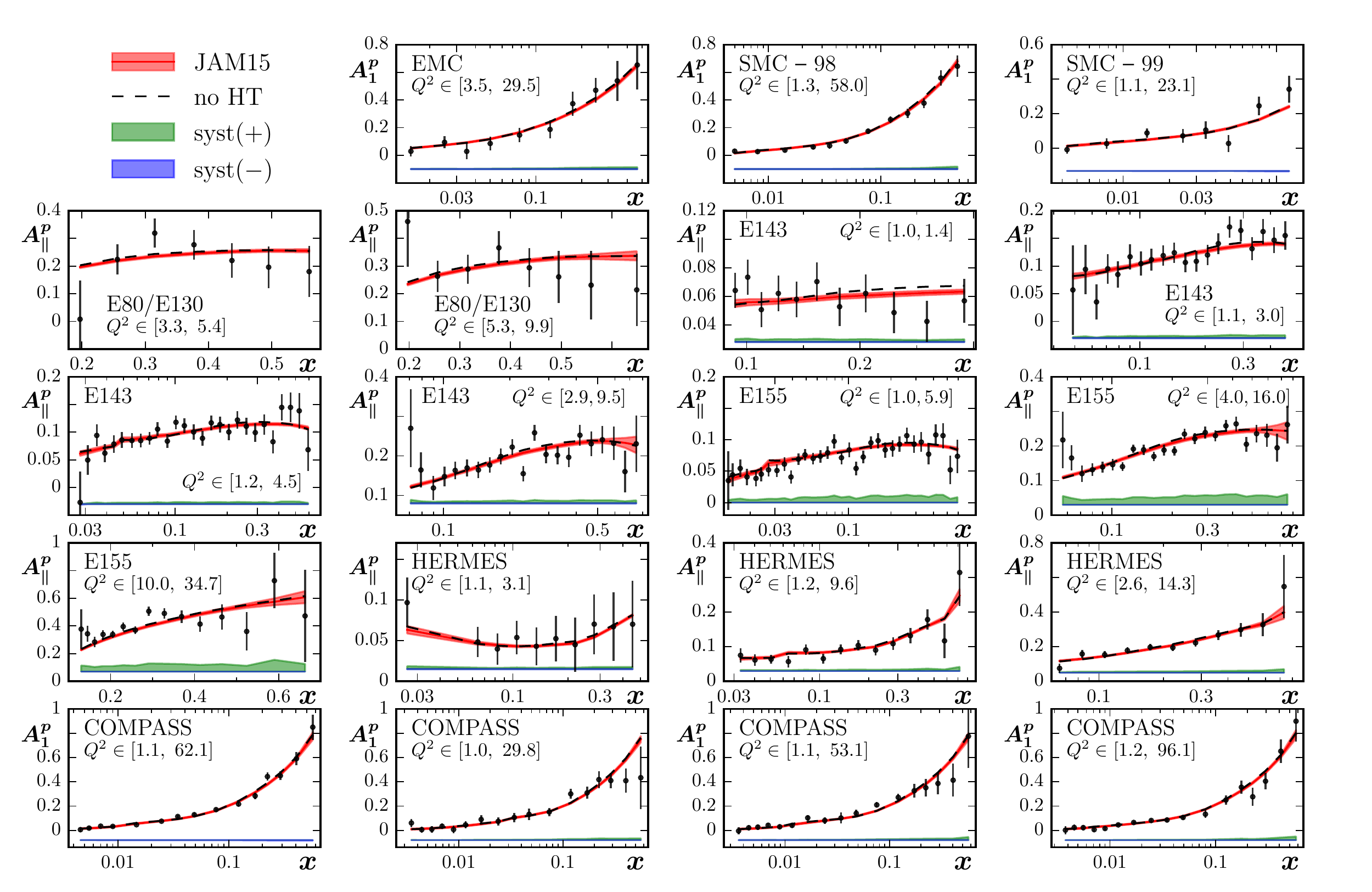}
\caption{Proton longitudinal polarization asymmetries
	$A_\|^p$ and $A_1^p$ from
	EMC \cite{EMC89},
	SMC \cite{SMC98, SMC99},
	COMPASS \cite{COMPASS10, COMPASS16},
	SLAC \cite{SLAC-E130, SLAC-E143, SLAC-E155p} and
	HERMES \cite{HERMES07} experiments.
	The $Q^2$ range	(in units of GeV$^2$) for the data in
	each panel is indicated.  The data are compared with the
	asymmetries from the JAM15 fit (solid red curves with bands
	indicating 1$\sigma$ uncertainties) and the contributions
	excluding higher twists (HT) (black dashed curves).
	The experimental data points include the normalization
	factors, and the systematic error bands indicate the
	positive (upper green [``syst($+$)''] bands) or
	negative (lower blue [``syst($-$)''] bands) shifts of the
	data from their nominal values.  Panels without visible
	systematic shifts correspond to data sets for which
	correlated uncertainties were not provided.}
\label{f.Apa_proton}
\end{figure}

\begin{figure}[t]	
\centering 
\includegraphics[width=1.05\textwidth]{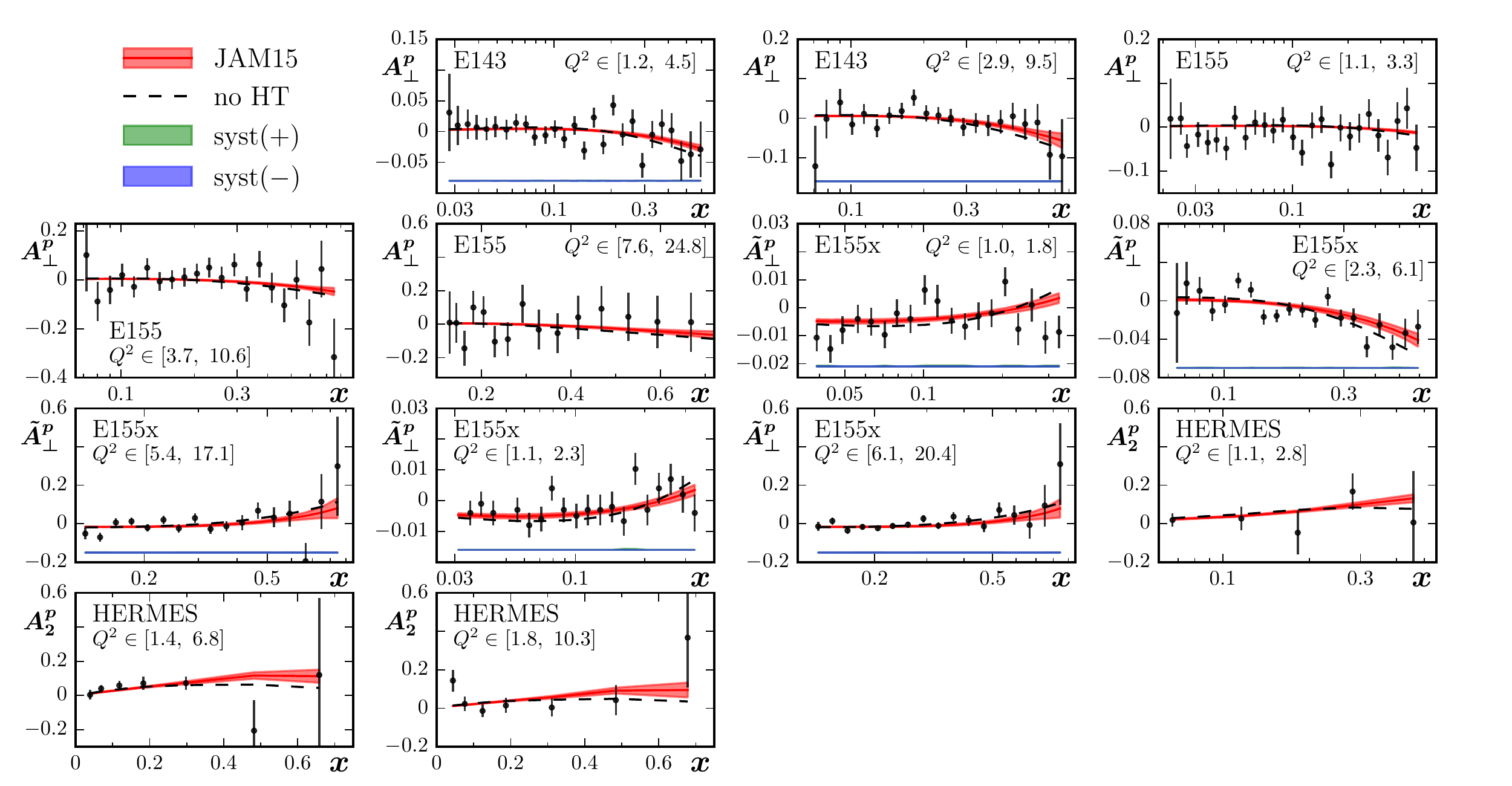}
\caption{Proton transverse polarization asymmetries
	$A_\perp^p$ and $A_2^p$ from
	SLAC \cite{SLAC-E143, SLAC-E155_A2pd, SLAC-E155x} and
	HERMES \cite{HERMES12}.
	The curves and legends are as in Fig.~\ref{f.Apa_proton}.}
\label{f.Ape_proton}
\end{figure}

\begin{figure}[t]	
\centering 
\includegraphics[width=1.05\textwidth]{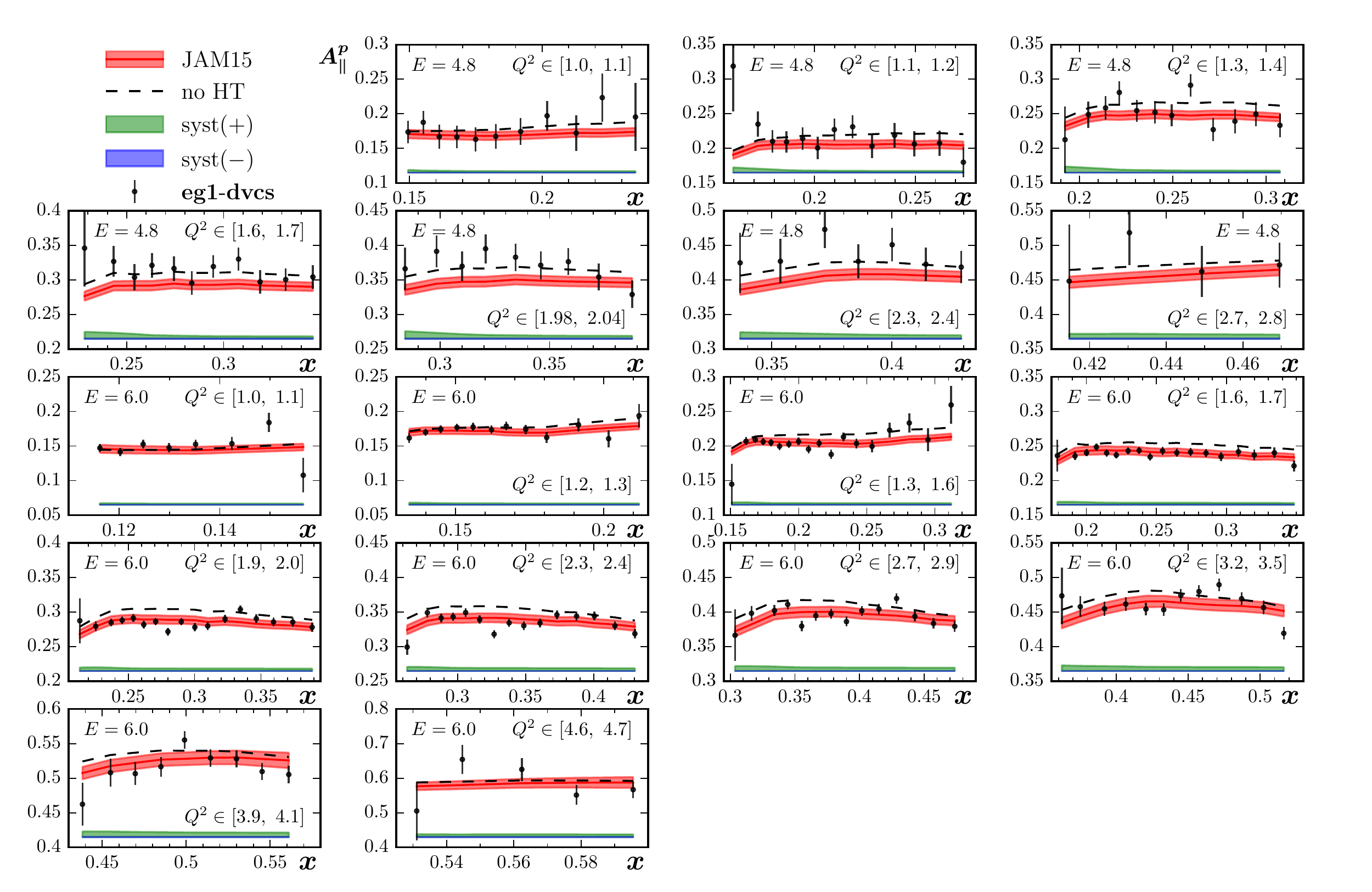}
\caption{Proton longitudinal polarization asymmetries $A_\|^p$
	from the eg1-dvcs \cite{eg1-dvcs} experiment at Jefferson Lab.
	The energies $E$ (in GeV) and $Q^2$ ranges (in GeV$^2$)
	for each panel are indicated.  The curves and legends are
	as in Fig.~\ref{f.Apa_proton}.}
\label{f.Apa_proton_dvcs}
\end{figure}

\begin{figure}[t]	
\centering 
\includegraphics[width=1.05\textwidth]{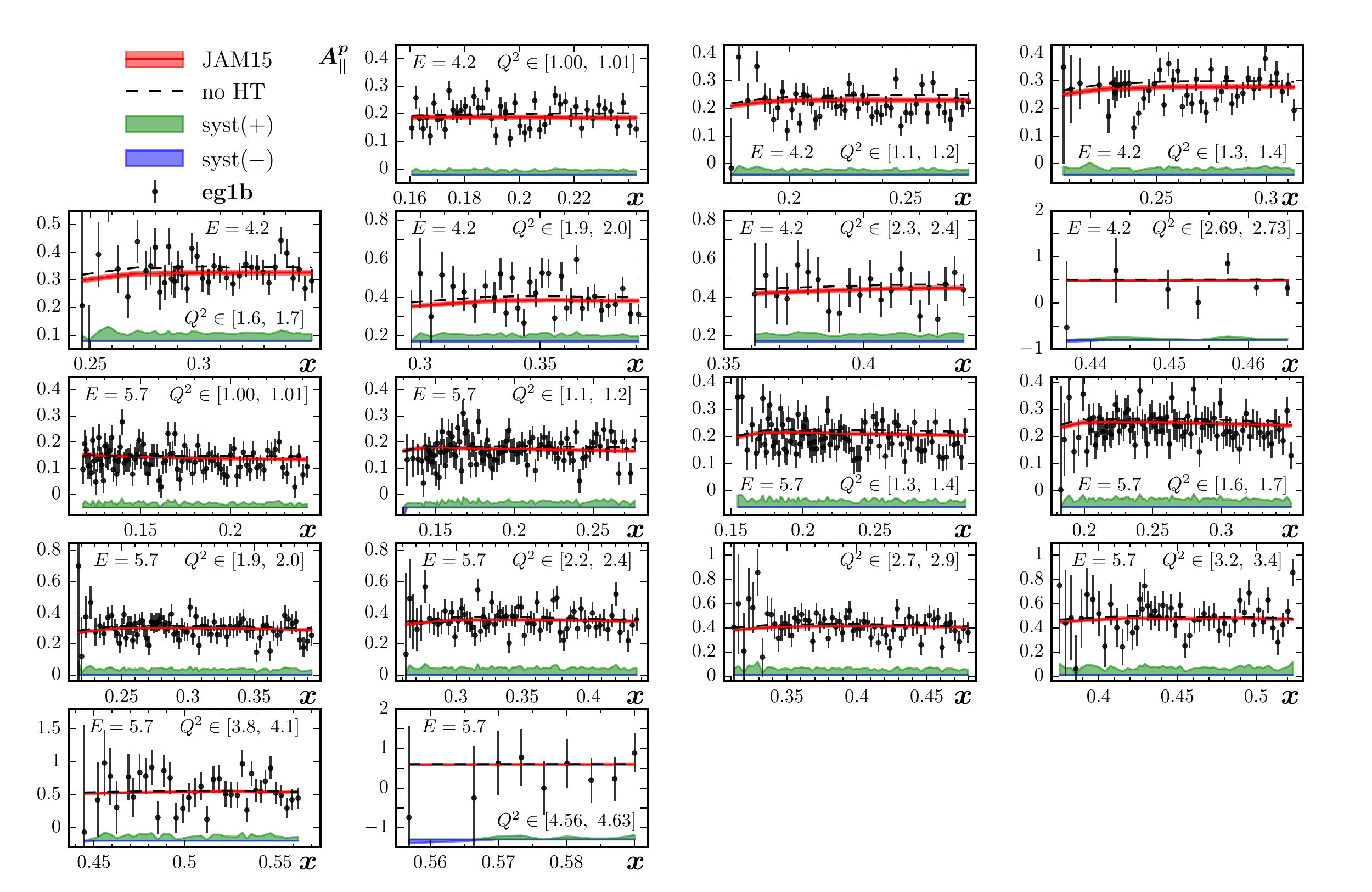}
\caption{Proton longitudinal polarization asymmetries $A_\|^p$
	from the eg1b \cite{eg1b-p} experiment at Jefferson Lab.
	The curves and legends are as in
	Fig.~\ref{f.Apa_proton_dvcs}.}
\label{f.Apa_proton_eg1b}
\end{figure}

\begin{figure}[t]	
\centering 
\includegraphics[width=1.05\textwidth]{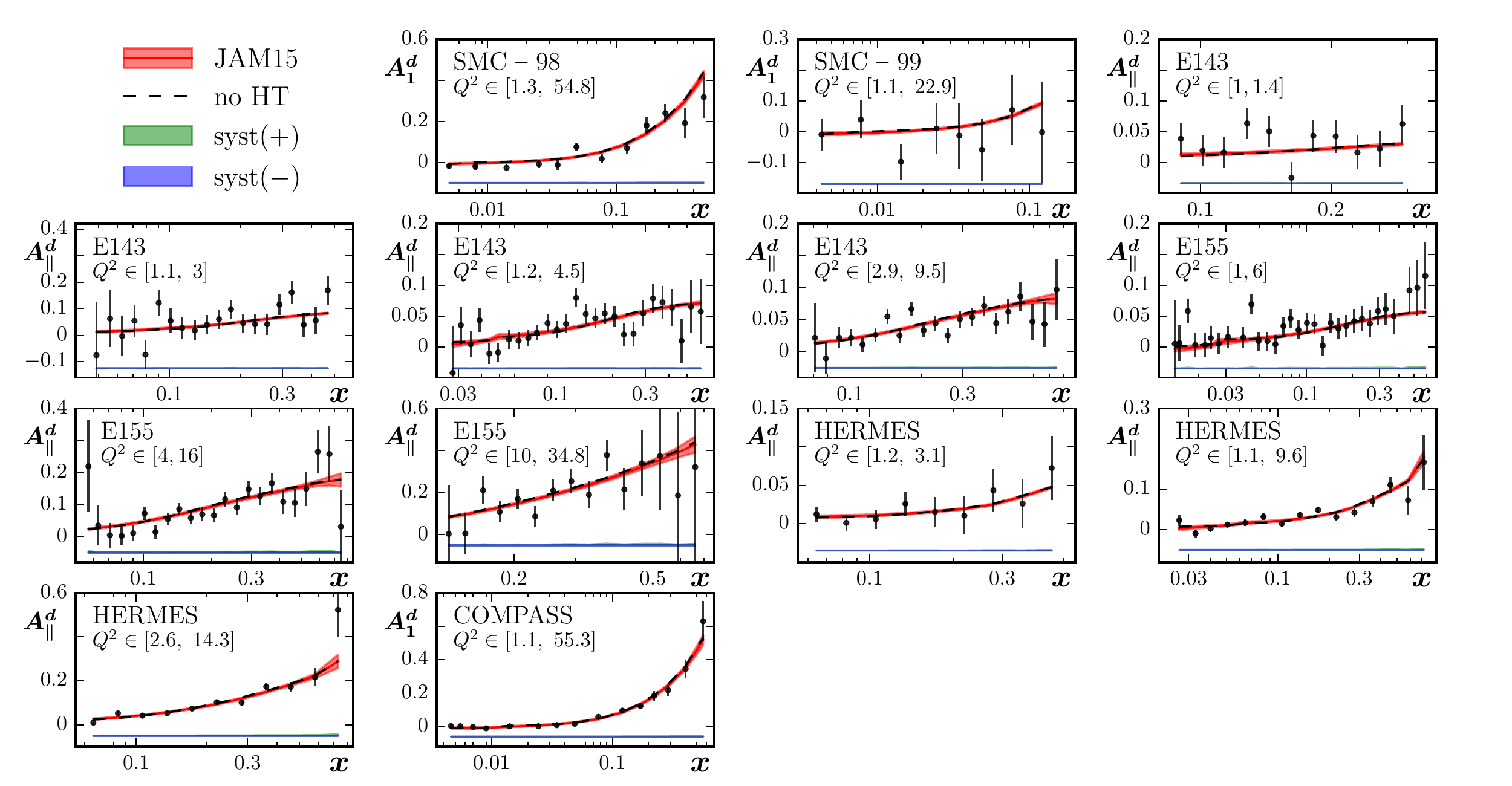}
\caption{Deuteron longitudinal polarization asymmetries
	$A_\|^d$ and $A_1^d$ from
	SMC \cite{SMC98, SMC99},
	COMPASS \cite{COMPASS10},
	SLAC \cite{SLAC-E143, SLAC-E155d} and
	HERMES \cite{HERMES07} experiments.
	The curves and legends are as in Fig.~\ref{f.Apa_proton}.}
\label{f.Apa_deuteron}
\end{figure}

\begin{figure}[t]	
\centering 
\includegraphics[width=1.05\textwidth]{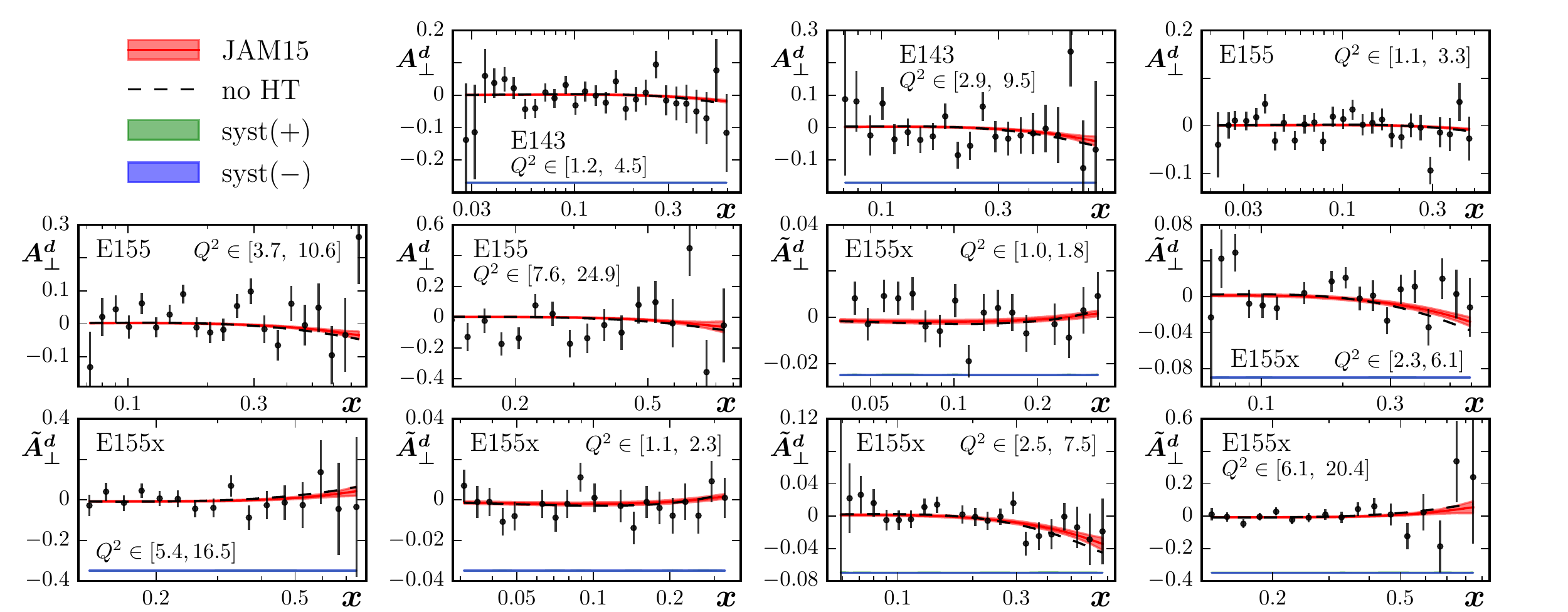}
\caption{Deuteron transverse polarization asymmetries
	$A_\perp^p$ from
	SLAC \cite{SLAC-E143, SLAC-E155_A2pd, SLAC-E155x} data.
	The curves and legends are as in Fig.~\ref{f.Apa_proton}.}
\label{f.Ape_deuteron}
\end{figure}

\begin{figure}[t]	
\centering 
\includegraphics[width=1.05\textwidth]{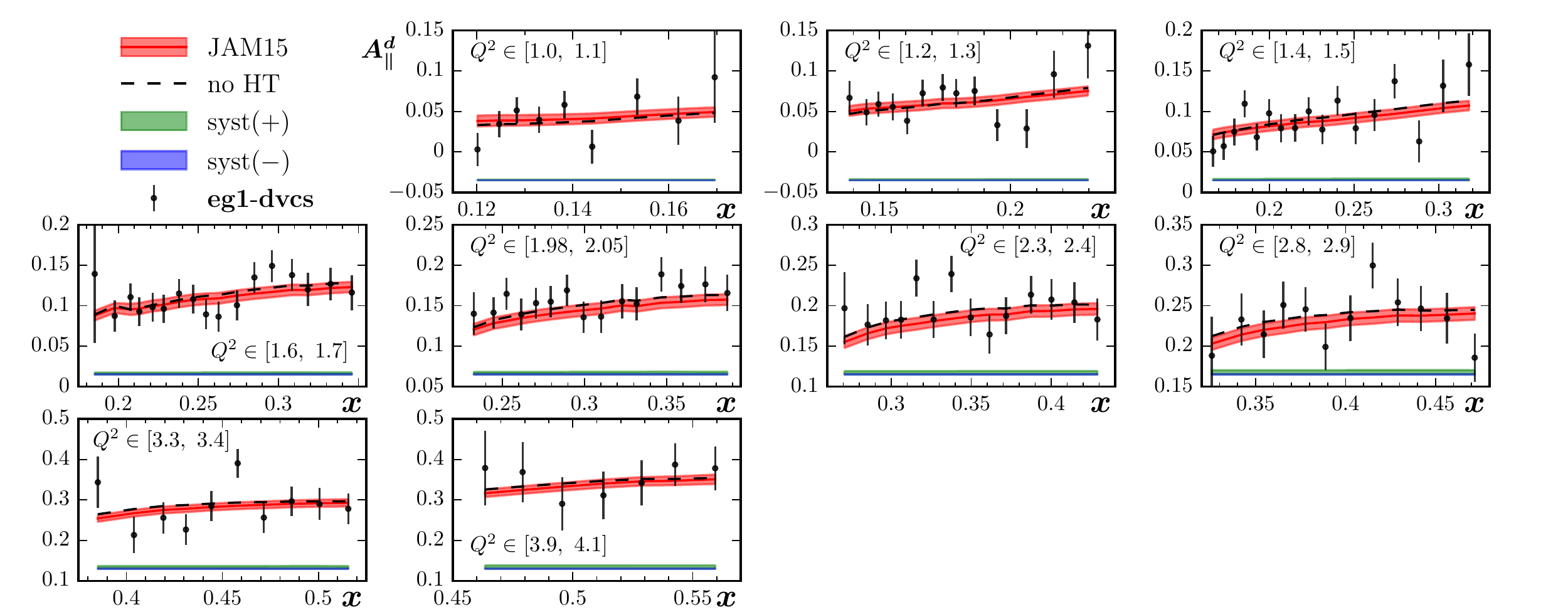}
\caption{Deuteron longitudinal polarization asymmetries $A_\|^d$
	from the eg1-dvcs \cite{eg1-dvcs} experiment at
	Jefferson Lab's Hall~B.  The curves and legends are
	as in Fig.~\ref{f.Apa_proton_dvcs}.}
\label{f.Apa_deuteron_dvcs}
\end{figure}

\begin{figure}[t]	
\centering 
\includegraphics[width=1.05\textwidth]{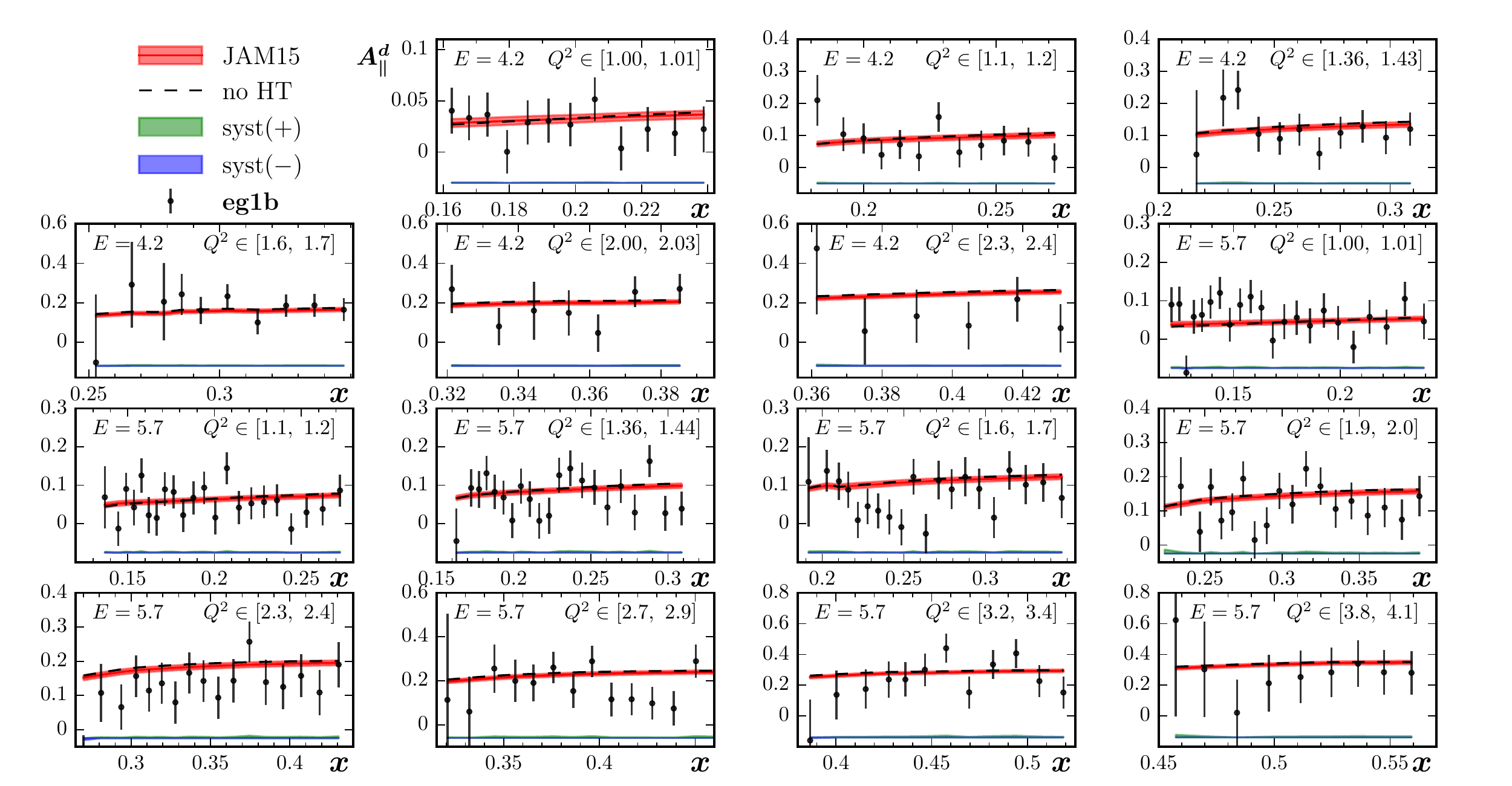}
\caption{Deuteron longitudinal polarization asymmetries $A_\|^d$
	from the eg1b \cite{eg1b-d} experiment at Jefferson Lab.
	The curves and legends are as in Fig.~\ref{f.Apa_proton_dvcs}.}
\label{f.Apa_deuteron_eg1b}
\end{figure}

\begin{figure}[t]	
\centering 
\includegraphics[width=1.05\textwidth]{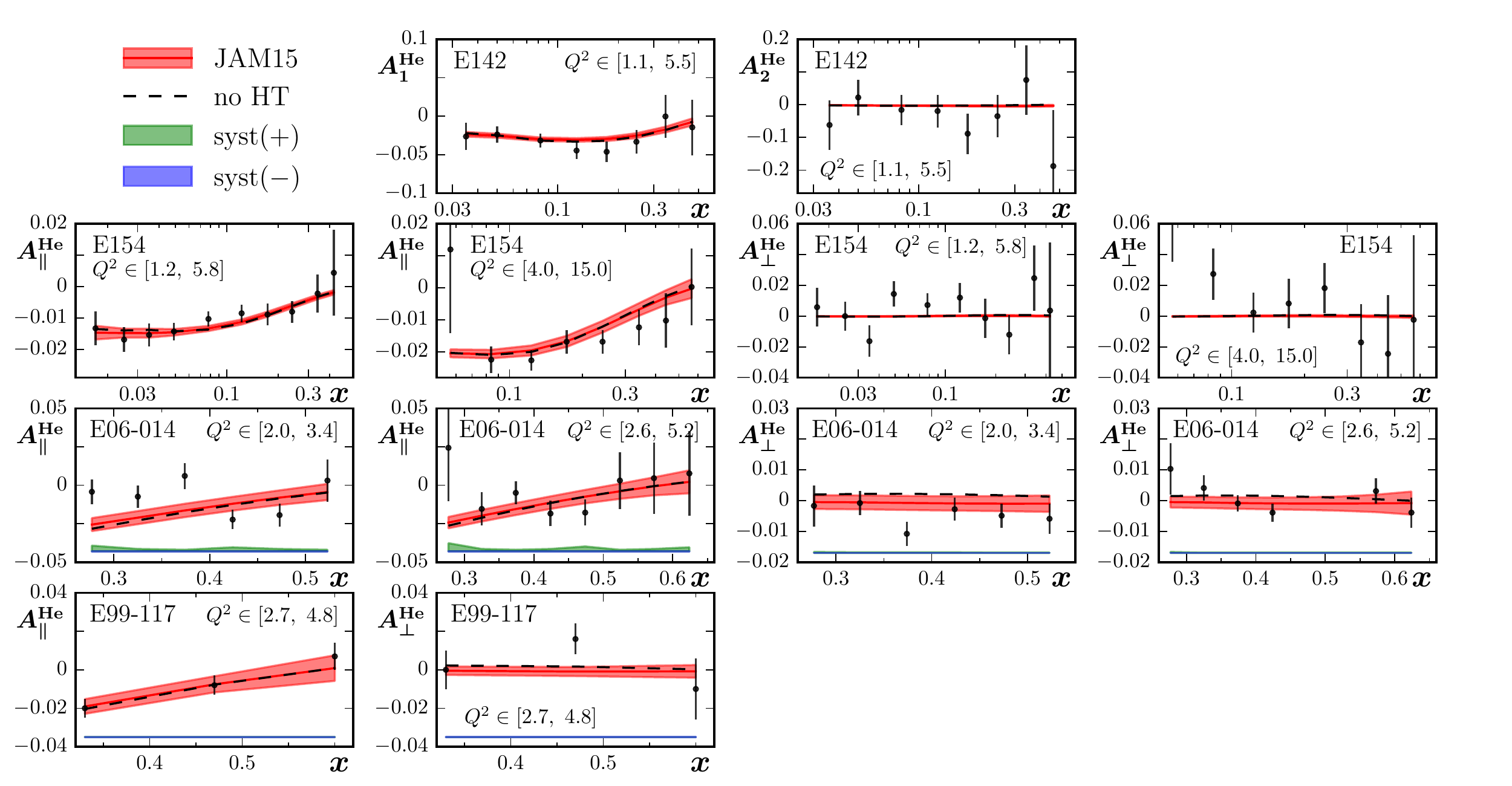}
\caption{$^3\rm He$ longitudinal ($A_\|^{\rm He}$, $A_1^{\rm He}$)
	and transverse ($A_\perp^{\rm He}$, $A_2^{\rm He}$)
	polarization asymmetries from
	SLAC \cite{SLAC-E142, SLAC-E154} and
	Jefferson Lab \cite{E99-117, E06-014_A1, E06-014_d2}
	experiments, compared with the JAM15 global fit.
	The curves and legends are as in Fig.~\ref{f.Apa_proton}.}
\label{f.helium} 
\end{figure}

The $\chisdof$ values for the individual data sets fitted in the JAM15
analysis are listed in Table~\ref{t.chi2}.  The overall $\chisdof$ is
1.07 for the 2515 data points in the global data set.  The fits to
the complete set of asymmetries used in analysis are illustrated in
Figs.~\ref{f.Apa_proton}--\ref{f.helium}.
In particular, the proton longitudinal polarization asymmetries
$A_\|^p$ and $A_1^p$ from the
EMC \cite{EMC89},			
SMC \cite{SMC98, SMC99},
COMPASS \cite{COMPASS10, COMPASS16},
SLAC \cite{SLAC-E130, SLAC-E143, SLAC-E155p}
and HERMES \cite{HERMES07}
experiments are shown in Fig.~\ref{f.Apa_proton} as a function
of $x$, for the various $Q^2$ ranges measured in the experiments,
ranging from $Q^2=1$~GeV$^2$ to $\sim 100$~GeV$^2$.
In each panel the measured asymmetries are compared with the
central values and uncertainties for the JAM15 fits, along with
the contributions to the asymmetries from leading twist only
(which include TMCs but not the higher twist terms).
The agreement between the JAM15 fit and the data is generally very
good over the entire range of $x$ and $Q^2$ spanned by these data,
and, with the exception of the most recent SMC \cite{SMC99} and
SLAC E155 \cite{SLAC-E155p} data, the $\chisdof$ values for each
experiment are less than one.

The error bars on each of the data points represent uncorrelated
uncertainties, while the upward or downward shifts of the data 
points due to their correlated uncertainties are indicated by
the upper (green) and lower (blue) bands, denoted by ``syst($+$)''
and ``syst($-$)'', respectively.  As discussed in
Sec.~\ref{ss.parametrization}, these shifts are computed by
fitting the point-by-point normalization factors $N_i^{(e)}$
in Eq.~(\ref{e.norm}) for each experimental data set.
The central values of the data points shown in
Figs.~\ref{f.Apa_proton}--\ref{f.helium} are then computed as
\begin{align}
\widetilde{{\cal D}}^{(e)}_i = N_i^{(e)} {\cal D}_i^{(e)},
\label{e.shifted_data}
\end{align}
and the uncorrelated uncertainties are given by
\begin{align}
\widetilde{\alpha}^{(e)}_i = N_i^{(e)} \alpha^{(e)}_i.
\label{e.shifted_data_err}
\end{align}
The systematic shifts syst($\pm$) are computed as the difference
$\widetilde{{\cal D}}^{(e)}_i - {\cal D}_i^{(e)}$ of the data points
from their nominal values.

The data on the proton transverse polarization asymmetries
$A_\perp^p$ and $A_2^p$ from the SLAC \cite{SLAC-E143,
SLAC-E155_A2pd, SLAC-E155x} and HERMES \cite{HERMES12}
experiments are compared in Fig.~\ref{f.Ape_proton}
with the JAM15 results.
The transverse asymmetries are generally very small, which
requires high precision experiments to extract nonzero values.
The agreement between the fit and the data is very good overall,
with $\chisdof \sim 1$ for all experiments other than SLAC E155x
\cite{SLAC-E155x}, where $\chisdof = 1.27$.
For both the longitudinal and transverse asymmetries, the
differences between the full JAM15 fit results and the 
leading twist contributions are very small.  There is an
indication of a slightly negative higher twist contribution
in the $A_\|^p$ data at $x \approx 0.2-0.4$ for
$Q^2 \lesssim 1.5$~GeV$^2$ in the SLAC E155 data \cite{SLAC-E155p},
and a slightly positive higher twist in the $A_\perp^p$ data at
larger $x$ values.

The effects of higher twists are more evident in the new
Jefferson Lab data in Figs.~\ref{f.Apa_proton_dvcs} and
\ref{f.Apa_proton_eg1b}, where the longitudinal proton
asymmetries $A_\|^p$ from the CLAS eg1-dvcs \cite{eg1-dvcs}
and eg1b \cite{eg1b-p} experiments, respectively,
are compared with the JAM15 fit.
(No Jefferson Lab transverse polarization data currently
exist for the proton, but will be available soon from
the SANE experiment in Hall~C \cite{SANE14}.)
The higher twists are generally negative and lead
to a decrease in $A_\|^p$ at the larger $x$ values
($x \gtrsim 0.2$) and low $Q^2 \lesssim 2$~GeV$^2$.
For the eg1b proton data, the fit to the nearly 900 points,
in fine bins of $x$ and $Q^2$, gives $\chisdof = 1.11$,
indicating relatively good agreement with both the lower
energy $E=4.2$~GeV and higher energy $E=5.7$~GeV data.
In some of the eg1b spectra (for example, in the $E=4.2$~GeV,
$Q^2 \in [1.3, 1.4]$~GeV$^2$ panel) there appear to be strong
correlations among the data, although these do not significantly
affect the overall $\chisdof$.

The eg1-dvcs data, on the other hand, have extremely small
statistical uncertainties and are more difficult to accommodate
within the global fit, as evidenced by the overall $\chisdof = 1.52$
for this data set.  This suggests that the uncorrelated uncertainties
here may be underestimated, particularly for the $E=6$~GeV data.
The very small errors on this data set dominate the $\chi^2$
fit to the Jefferson Lab data, and lead to an upward systematic
pull on the eg1b data, as indicated by the predominantly
syst($+$) band for the correlated uncertainties.
A comparison of the entire eg1-dvcs data set reveals the
existence of a possible tension between the $E=4.8$~GeV and
6~GeV data, with the fitted results lying systematically below
the lower-energy data for $Q^2 \approx 1.5-2.5$~GeV$^2$.
Large systematic shifts of the data relative to the JAM15 fit
are less evident for the $E=6$~GeV data because the smaller
uncertainties here provide a stronger pull on the fit.

Similar features are seen in the deuteron longitudinal and
transverse asymmetry data, illustrated in Figs.~\ref{f.Apa_deuteron}
and \ref{f.Ape_deuteron}, respectively, for the earlier measurements
from
SMC \cite{SMC98, SMC99},
COMPASS \cite{COMPASS10},
SLAC \cite{SLAC-E143, SLAC-E155d, SLAC-E155_A2pd, SLAC-E155x} and
HERMES \cite{HERMES07}.
Generally the deuteron asymmetry data have larger uncertainties
compared with the proton data.
Most of the data sets can be well described by the global fit,
with only the SMC $A_1^d$ data \cite{SMC98} and E155 $A_\perp^d$
data \cite{SLAC-E155d} having moderately large $\chisdof$ values
(1.26 and 1.52, respectively).  The former comes mostly from the
small errors on the low-$x$ data, while the scatter of the points
in the latter, especially at the higher $Q^2$ values, suggests
a possible underestimation of uncorrelated uncertainties.
For the longitudinal asymmetries $A_\|^d$ and $A_1^d$ the differences
between the full JAM15 results and the leading twist contributions are
negligible.  For the transverse polarization asymmetries $A_\perp^p$
there is a slight indication of nonzero higher twists at the highest
$x$ values, but the effects are very small on the scale of the
experimental uncertainties.

The more recent deuteron $A_\|^d$ data from the Jefferson Lab
eg1-dvcs \cite{eg1-dvcs} and eg1b \cite{eg1b-d} experiments are shown
in Figs.~\ref{f.Apa_deuteron_dvcs} and \ref{f.Apa_deuteron_eg1b},
respectively, compared with the JAM15 fit.
Good fits with $\chisdof \approx 1$ are found for both the eg1-dvcs
and eg1b data sets.  The similarity between the full results and the
leading twist contributions indicates no significant higher twists
within the experimental uncertainties.
The systematic shifts syst($\pm$) for the deuteron data are much
smaller than for the corresponding proton asymmetries, mostly
because of the somewhat larger uncorrelated uncertainties.
For the eg1b data there is a small tendency for the global fit
to overestimate the experimental asymmetries, especially for the
$E=5.7$~GeV energy data.

Finally, the world's data on longitudinal and transverse
polarization asymmetries of $^3\rm He$ are displayed in
Fig.~\ref{f.helium} for the SLAC E142 \cite{SLAC-E142} and
E154 \cite{SLAC-E154} experiments, and the \mbox{E99-117}
\cite{E99-117} and \mbox{E06-014} \cite{E06-014_A1, E06-014_d2}
experiments in Jefferson Lab's Hall~A.  As in the case of the
deuteron data, there is no evidence for large higher twists in
the $A_\|^{\rm He}$ asymmetries, but there is an indication
of a small negative higher twist contribution to $A_\perp^{\rm He}$
in the \mbox{E06-014} data at the lower $Q^2$ values.
Generally the fits give small $\chisdof$ values for all the
longitudinal asymmetry data sets, with the exception of the
\mbox{E06-014} $A_\|^{\rm He}$ data set which has $\chisdof = 2.12$.
Comparison with the JAM15 fit here suggests an incompatibility
with the data at the smaller $x$ values.
Similarly, good fits are also obtained for the transverse
polarization data, with a large $\chisdof$ ($\gtrsim 1.5$)
observed only for the \mbox{E99-117} $A_\perp^{\rm He}$ data.
However, this comes mostly from a single datum, and because
the data set contains a total of only 3 points.

\subsection{Impact of JLab data}
\label{ss.jlab}

To assess more quantitatively the impact of the new Jefferson Lab
data on the global fit, we perform an independent IMC analysis of
the world's data without inclusion of any of the measurements from
Refs.~\cite{E99-117, E06-014_A1, E06-014_d2, eg1-dvcs, eg1b-p, eg1b-d}.
The results of the IMC fits with and without the Jefferson Lab data
are presented in \fref{impact} for the \mbox{twist-2} $\Delta u^+$,
$\Delta d^+$, $\Delta s^+$ and $\Delta g$ PDFs, the \mbox{twist-3}
$D_u$ and $D_d$ PDFs, and the \mbox{twist-4} proton and neutron
distributions $H_p$ and $H_n$, as a function of $x$ at $Q^2=1$~GeV$^2$.
Although the complete IMC analysis contains around 8000 fits, for
clarity in \fref{impact} we illustrate the results by a random
sample of 50 fits.

\begin{figure}[t]
\includegraphics[width=\textwidth]{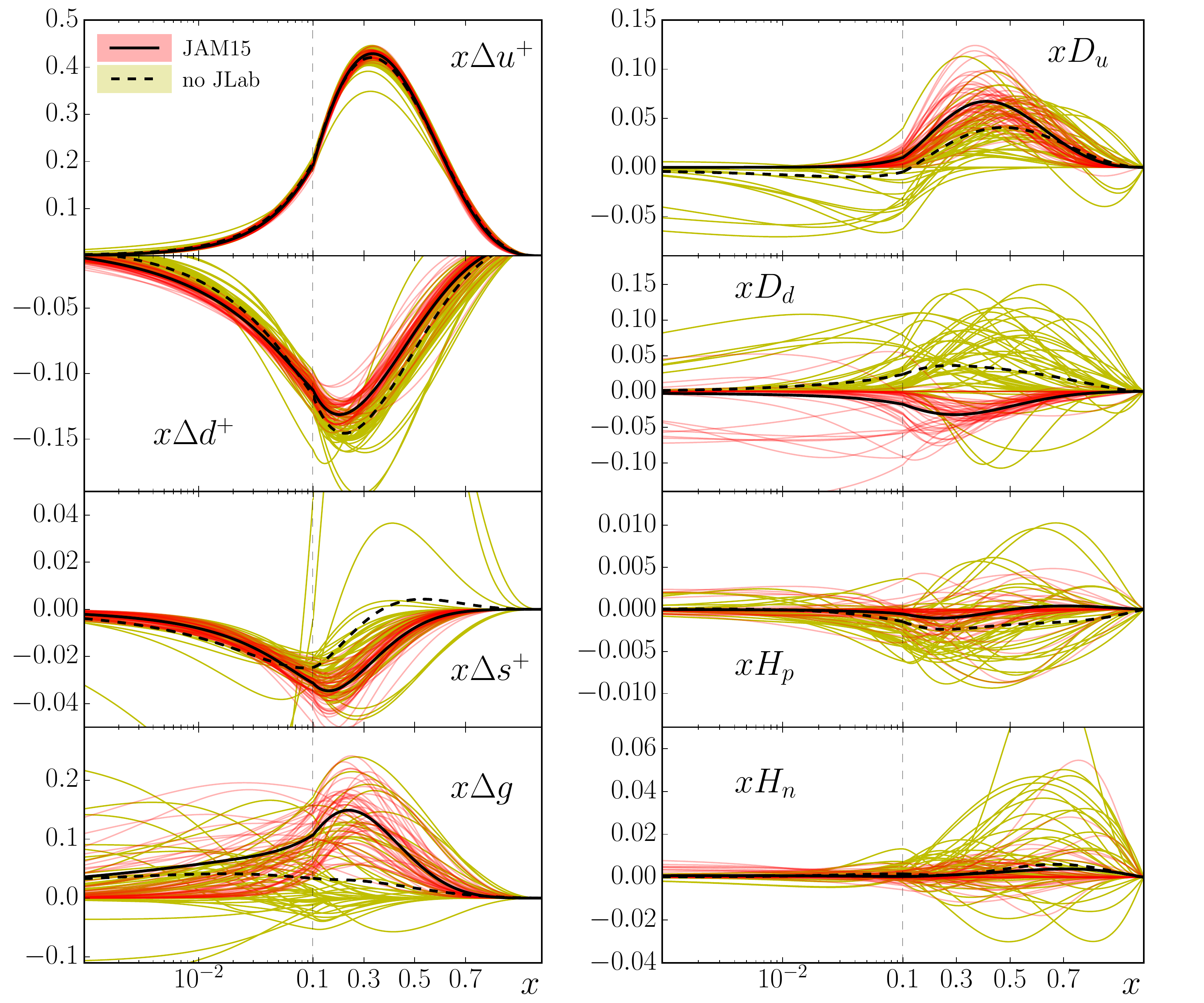} 
\caption{Comparison of the JAM15 IMC fits (red curves, with
	the average indicated by the black solid curve) with
	corresponding fits excluding all Jefferson Lab data
	(yellow curves, with the average given by the black
	dashed curve) for the twist-2 PDFs $\Delta u^+$,
	$\Delta d^+$, $\Delta s^+$ and $\Delta g$,
	the twist-3 distributions $D_u$ and $D_d$,
	and the twist-4 functions $H_p$ and $H_n$ at
	$Q^2=1$~GeV$^2$.  Note that $x$ times the distribution
	is shown.  For illustration each distribution is
	represented by a random sample of 50 fits.}
\label{f.impact} 
\end{figure}

The inclusion of the Jefferson Lab data results in a reduction
of the uncertainty bands on the $\Delta u^+$ and $\Delta d^+$ PDFs
in the region $0.1 \lesssim x \lesssim 0.7$ where the Jefferson Lab
data are localized.  This may be expected given that these
distributions give the leading contributions to the inclusive
DIS asymmetries at these kinematics.

Interestingly, however, we also observe significant reduction of the
uncertainties in $\Delta u^+$ and $\Delta d^+$ at small values of $x$,
outside of the kinematic range of the Jefferson Lab experiments.
By studying the correlations between PDFs over the entire $x$ range,
which are partly induced by the weak baryon decay constraints
[Eqs.~(\ref{e.gA}) and (\ref{e.a8})], we find a strong anticorrelation
between the $\Delta u^+$ distribution at large and small $x$ values.
Since the Jefferson Lab data tend to favor a higher $\Delta u^+$
in the region $0.1 \lesssim x \lesssim 0.7$, the anticorrelation
has the effect of favoring a suppressed $\Delta u^+$ at low $x$.
Similar arguments hold also for $\Delta d^+$ PDF.

In the absence of Jefferson Lab data, a strong correlation also
exists between higher values of the polarized strange PDF
$\Delta s^+$ at $x \sim 0.4$ and higher $\Delta u^+$ at small $x$.
The disfavoring by the data of the latter then indirectly constrains
the strange distribution to have smaller values across all $x$.
The uncertainty on $\Delta s^+$ is also significantly larger
without the Jefferson Lab constraints, as indicated by the
larger spread of the fitted results in Fig.~\ref{f.impact}.
The strange quark distribution illustrates the point that in the
Monte Carlo approach there is no guarantee that the final posteriors
will be clustered in a specific region of parameter space.  For
example, two distinct solutions can describe the same PDF in some
neighborhood of $x$, while deviating in other $x$ regions; data cannot
distinguish the two solutions due to correlations.  Such a picture of
multiple regions and error bands is absent in traditional single-fit
analyses, where the effect of adding more data means that the $\chis$
is steeper around the minimum.  While this is also true for Monte Carlo
fits, in the IMC approach, however, the error bands in practice cover
more than one minimum, if multiple solutions are present.

The $\Delta s^+$ PDF is also indirectly impacted by the different
$Q^2$ evolution of the singlet and nonsinglet distributions,
especially with the greater statistics at lower $Q^2$ values
afforded by the Jefferson Lab data.
The $Q^2$ evolution also provides a way of indirectly constraining
the polarized gluon distribution $\Delta g$, in the absence of
jet data from polarized $pp$ collisions \cite{STAR_jet15} in the
current analysis.
Indeed, as Fig.~\ref{f.impact} indicates, the new Jefferson Lab
results actually prefer a more positive $\Delta g$ distribution at
intermediate $x$ values, $x \approx 0.1-0.5$, with a smaller spread of
possible behaviors, but with still large uncertainties at lower $x$.

In the higher twist sector, as one might expect, the greater
abundance of lower-$Q^2$ data provides even more stringent
constraints on the twist-3 and twist-4 distributions.
In particular, the global analysis reveals that with the
addition of Jefferson Lab data the twist-3 $D_u$ distribution
becomes more positive at $x > 0.1$, while the $D_d$ distribution
effectively switches sign to become negative and smaller in
magnitude.  The twist-3 distributions thus acquire the same
signs for the $u$ and $d$ flavors as their twist-2 PDF analogs.

For the twist-4 distributions, while $H_p$ and $H_n$ are largely
unconstrained in the fit without Jefferson Lab data, in the full
fit the spread is reduced considerably, and the results for both
distributions are consistent with zero.
The dominant contributions of the higher twists to the DIS
asymmetries are therefore driven by the twist-3 terms.

\subsection{JAM15 distributions and moments}
\label{ss.PDFs}

\begin{figure}[t]
\centering 
\includegraphics[width=\textwidth]{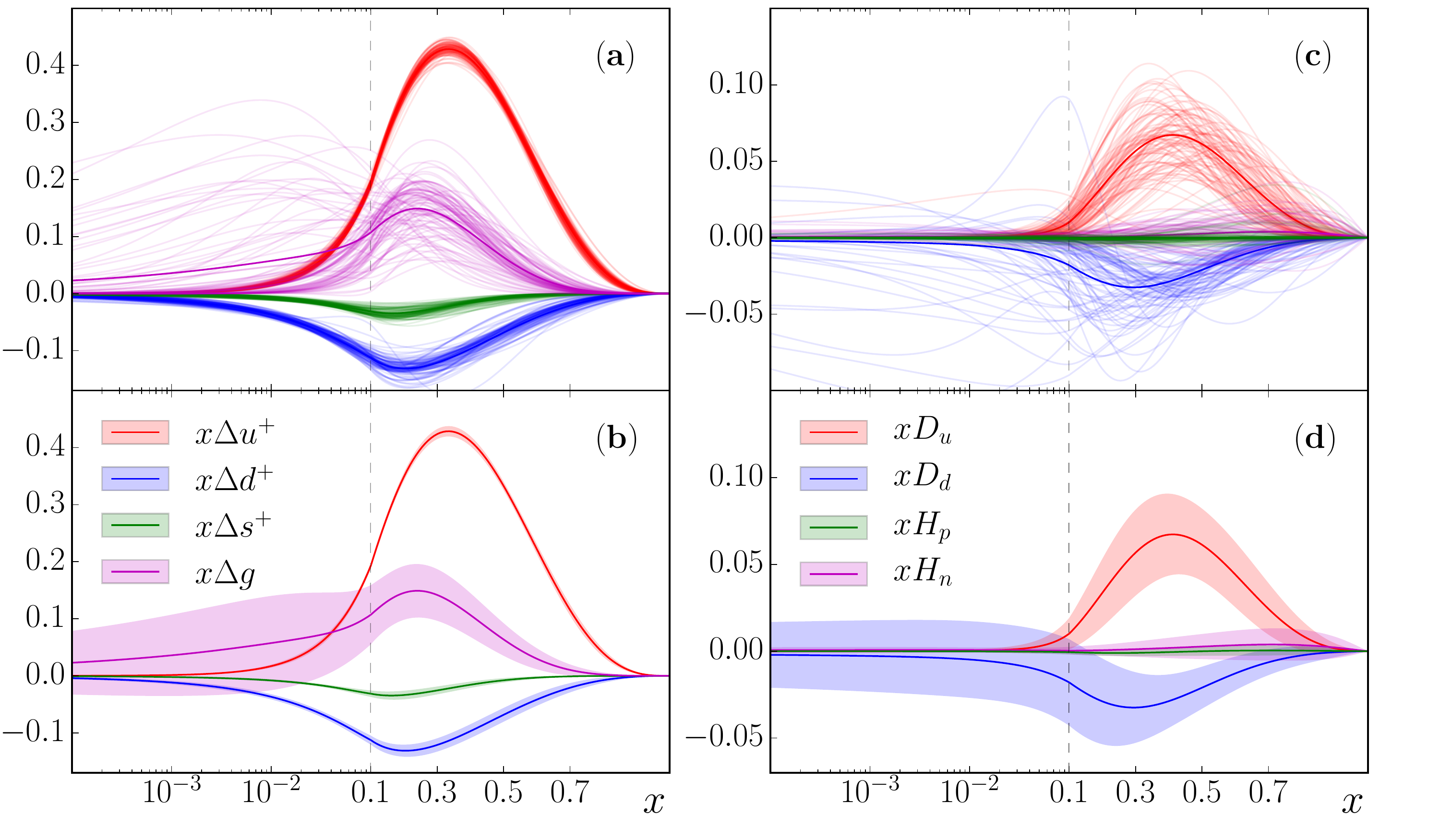}
\caption{Leading twist $\Delta u^+$, $\Delta d^+$, $\Delta s^+$
	and $\Delta g$ distributions [{\bf (a)} and {\bf (b)}]
	and the	higher twist $D_{u, d}$ and $H_{p, n}$ distributions
	[{\bf (c)} and {\bf (d)}] as a function of $x$ for
	$Q^2=1$~GeV$^2$.
	Panels {\bf (a)} and {\bf (c)} show a random sample of
	100 from the 8000 IMC fits, while {\bf (b)} and {\bf (d)}
	show the average distributions and the standard deviations
	computed from Eqs.~(\ref{e.expectation}) and
	(\ref{e.variance}).  Note that $x$ times the distribution
	is shown.}
\label{f.LT_dist} 
\end{figure}

The final distributions for the full JAM15 fit are displayed in
Fig.~\ref{f.LT_dist} as a function of $x$ at fixed $Q^2=1$~GeV$^2$,
with the leading twist PDFs and the higher twist distributions
for different flavors shown on the same graph for comparison.
To illustrate the Monte Carlo aspect of the analysis, a random
selection of 100 fits from the full sample of $\approx 8000$ in
the full analysis is shown, along with the expectation values
and standard deviations for each distribution computed from
Eqs.~(\ref{e.expectation}) and (\ref{e.variance}) using the
full sample.
The $\Delta u^+$ and $\Delta d^+$ PDFs are the best determined
distributions from the inclusive DIS data, with relatively small
uncertainty bands.  We stress that the uncertainties here are
computed unambiguously from the Monte Carlo analysis, independent
of any tolerance criteria, which are sometimes invoked in single-fit
analyses to inflate PDF errors when fitting incompatible data sets
\cite{PJD13}.
Integrated over all $x$, the lowest moments of the
$\Delta u^+$ and $\Delta d^+$ distributions are
$0.83 \pm 0.01$ and $-0.42 \pm 0.01$, respectively.
The contributions from the extrapolated regions, $x < 0.001$
and $x > 0.8$, where the PDFs are not directly constrained by
data, are very small as a comparison between the truncated
and full moments in Table~\ref{t.moments} demonstrates.

\begin{table}[t!]
\caption{Lowest moments of the twist-2 PDFs $\Delta u^+$,
	$\Delta d^+$, $\Delta s^+$, $\Delta \Sigma$ and
	$\Delta G$, the twist-3 $d_2^p$ and $d_2^n$ moments,
	and the \mbox{$x^2$-weighted} moments $h_p$ and $h_n$
	of the twist-4 distributions.
	The truncated moments in the measured region
	$x \in [0.001,0.8]$ and the extrapolated full
	moments are shown at $Q^2=1$~GeV$^2$. \\}
\begin{tabular*}{0.7\columnwidth}{@{\extracolsep{\fill}} ccc} \hline\hline
~moment~       	 & truncated~~           & full~		\\ \hline
~$\Delta u^+$~~  & $ 0.82 \pm 0.01$~     & $ 0.83  \pm 0.01$~	\\ 
~$\Delta d^+$~~  & $-0.42 \pm 0.01$~~~   & $-0.44  \pm 0.01$~~~ \\
~$\Delta s^+$~~  & $-0.10 \pm 0.01$~~~   & $-0.10  \pm 0.01$~~~ \\
~$\Delta \Sigma$~~~ & $ 0.31 \pm 0.03$~  & $ 0.28  \pm 0.04$~	\\
~$\Delta G  $~~~ & $ 0.5  \pm 0.4$~      & $ ~~1   \pm 15$~	\\ \hline
~$d_2^p$~~~      & $ 0.005 \pm 0.002$~   & $ 0.005 \pm 0.002$~	\\
~$d_2^n$~~~      & $-0.001 \pm 0.001$~~~ & $-0.001 \pm 0.001$~~~ \\
~$h_p$~~~        & $-0.000 \pm 0.001$~~~ & $ 0.000 \pm 0.001$~	\\
~$h_n$~~~	 & $ 0.001 \pm 0.002$~   & $ 0.001 \pm 0.003$~	\\ \hline
\label{t.moments} 
\end{tabular*}
\end{table}

The strange quark distribution $\Delta s^+$ turns out to be negative,
constrained by a combination of $Q^2$ evolution, weak baryon decay
constants, and the assumption of an SU(3) symmetric sea,
Eq.~(\ref{e.SU3}).  The value of $\Delta s^+$ integrated over $x$
is $-0.10 \pm 0.01$, which then implies a total helicity carried by
quarks and antiquarks of $\Delta\Sigma = 0.28 \pm 0.04$ at the input
scale.  The extrapolated region contributes little to the moments of
the quark distributions, in contrast to the gluon case, where the
unmeasured region plays a much more important role.
In particular, while the gluon helicity from the experimentally
constrained region is $0.5 \pm 0.4$, the total moment approximately
doubles in magnitude, but with a significantly larger uncertainty,
$\Delta G = 1 \pm 15$.  This is reflected by the much wider error band
on the $\Delta g(x)$ distribution in Fig.~\ref{f.LT_dist} than on the
polarized quark PDFs.  The uncertainty is expected to be reduced once
jet and pion production data from polarized $pp$ collisions are
included in the analysis \cite{JAM16}.

\begin{figure}[t]
\includegraphics[width=\textwidth]{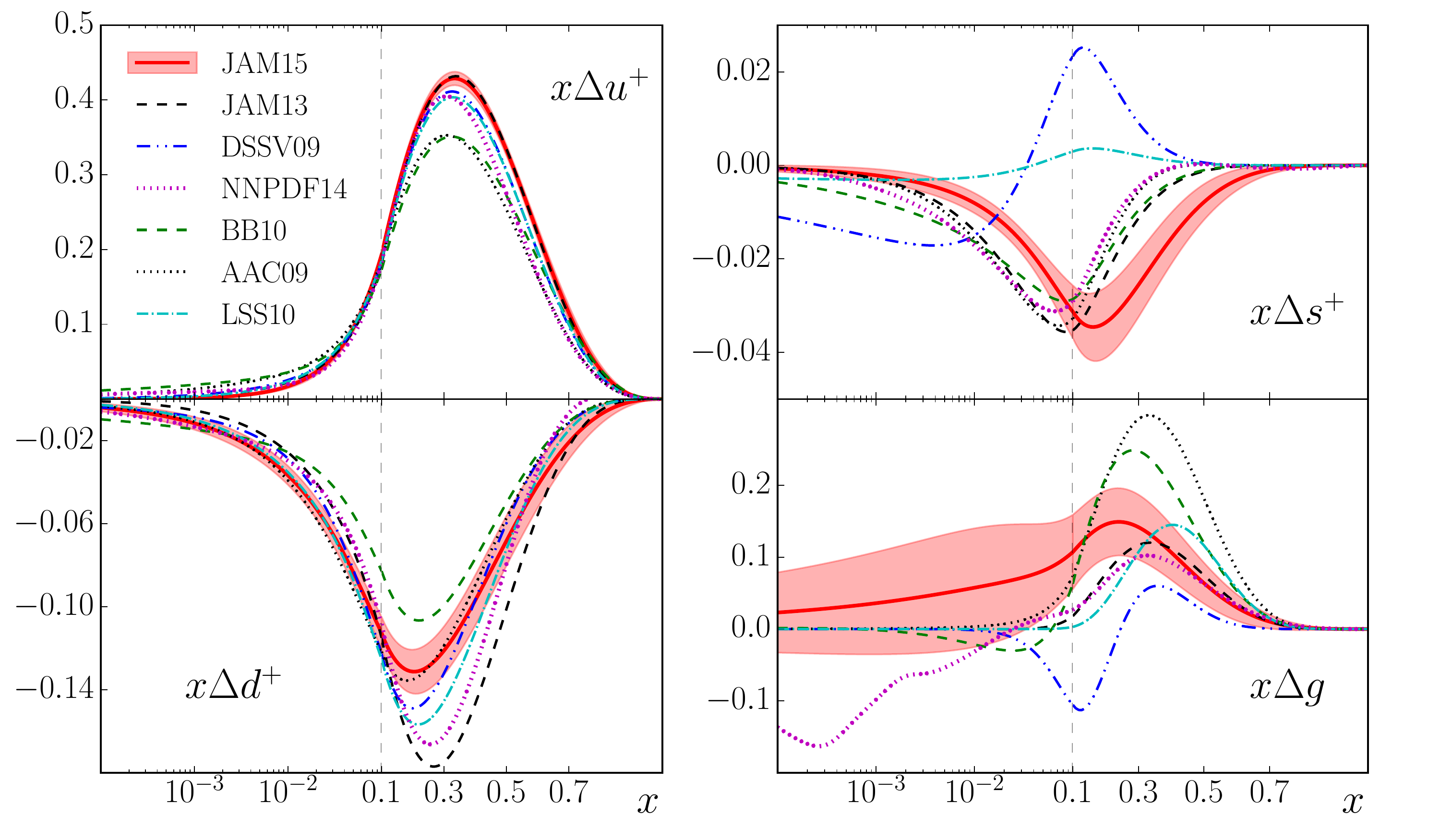} 
\caption{Comparison of the JAM15 PDFs $\Delta u^+$, $\Delta d^+$,
	$\Delta s^+$ and $\Delta g$ at $Q^2=1$~GeV$^2$ with
	PDFs from other parametrizations in the literature,
	including DSSV09 \cite{DSSV09}, NNPDF14 \cite{NNPDF14},
	BB10 \cite{BB10}, AAC09 \cite{AAC09}, LSS10 \cite{LSS10},
	and JAM13 \cite{JAM13}.}
\label{f.otherPDFs}
\end{figure}

The difficulty in constraining the polarized gluon distribution
is clearly revealed through the spread of $\Delta g$ from various
global PDF parametrizations illustrated in Fig.~\ref{f.otherPDFs}.
Here the PDFs from the
DSSV09 \cite{DSSV09},
AAC09 \cite{AAC09},
BB10 \cite{BB10},
LSS10 \cite{LSS10}
and NNPDF14 \cite{NNPDF14} global analyses are compared with
the JAM15 results, and with the previous JAM13 \cite{JAM13}
distributions.
Note that the BB10 fit uses only inclusive DIS data, similar
to our analysis and JAM13, while LSS10 includes also
semi-inclusive DIS asymmetries.  The other analyses consider
in addition data from polarized $pp$ scattering with jet and
$\pi$ production at RHIC, which have the strongest constraints
on the gluon polarization, while NNPDF14 also includes $W$ boson
asymmetries to constrain the antiquark sea.

In most of the fits the $\Delta g$ PDF is positive at large $x$,
with a sign change at smaller $x$ values for the DSSV09, BB10
and NNPDF14 PDFs.  Even though a node is allowed in the JAM15
parametrization, our analysis with inclusive DIS data only
does not favor a sign change.
Depending on which data sets are included in the fits, the
integrated gluon moment $\Delta G$ can vary enormously between
the parametrizations.
Interestingly, the latest analysis by de~Florian {\it et al.}
\cite{DSSV14} of the recent high-statistics jet data from RHIC
also gives a positive $\Delta g$ distribution, qualitatively
similar to the JAM15 result, with no indication of a sign change
in the measured $x$ region.

The sign of the $\Delta s^+$ distribution is consistent with that
found in previous global PDF analyses based on inclusive DIS data,
as Fig.~\ref{f.otherPDFs} illustrates.  As a function of $x$, the
shape of the JAM15 $\Delta s^+$ is slightly harder than for other PDF
parametrizations, which stems from the inclusion of the Jefferson Lab
Hall~B data \cite{eg1b-p, eg1-dvcs} and the correlations with the
polarized $u$ and $d$ distributions (see below).
A softer polarized strange distribution could be obtained by
enforcing a larger value for the $b$ parameter in
Eq.~(\ref{e.parametrization}), as is assumed in many of the
single-fit PDF analyses.
In our IMC analysis we allow the strange quark $b$ parameter
in the initial sampling to be as large as 10; however, the
Monte Carlo fits prefer smaller values.
%
%
In contrast to the negative $\Delta s^+$ obtained from the analysis
of DIS asymmetries, inclusion of the semi-inclusive kaon production
data in the DSSV09 and LSS10 fits induces a positive $\Delta s^+$
at $x \gtrsim 0.05$.
Currently the tension between the inclusive and semi-inclusive DIS data
and their impact on the sign of the polarized strange distribution is
not completely understood \cite{LSS11, LSS15}, and the definitive
extraction of $\Delta s^+$ will require careful treatment of all
processes to which strange quarks contribute, as well as a reliable
determination of fragmentation functions.

For the much better determined $\Delta u^+$ and $\Delta d^+$
distributions, the shapes and magnitudes from the JAM15 fit
are generally similar to those found in previous analyses,
but with some important features.
The $\Delta u^+$ PDF is slightly higher at intermediate
$x \approx 0.3-0.5$ than in most of the other analyses, as was
the case for the JAM13 distribution, but overall the spread
between the different parametrizations is relatively small.
The BB10 and AAC09 $\Delta u^+$ distributions have the smallest
magnitude at the peak, $\approx 20\%$ smaller than JAM15.

The $\Delta d^+$ distribution, on the other hand, is somewhat
less negative at $x \gtrsim 0.1$ than the JAM13 result, but
similar to the DSSV09 and AAC09 distributions.  Interestingly,
the JAM15 $\Delta d^+$ PDF is also similar to the ``reference''
fit from the JAM13 analysis \cite{JAM13}, which did not include
any nuclear smearing or finite-$Q^2$ corrections.  As shown in
Ref.~\cite{JAM13}, nuclear smearing and higher twist corrections
in particular render $\Delta d^+$ more negative for $x \gtrsim 0.2$.
Inclusion of the new Jefferson Lab data make $\Delta d^+$ less
negative, countering the effects of the nuclear and hadronic
corrections.
Because of the weak baryon constraints on the moments of the
quark PDFs, many aspects of the $\Delta u^+$, $\Delta d^+$
and $\Delta s^+$ distributions and their uncertainties are
strongly correlated.
Compared with the JAM13 distributions, for example, the shift
in the JAM15 $\Delta d^+$ PDF towards more positive values at
$x \gtrsim 0.2$ is directly correlated with the shift of the
$\Delta s^+$ toward more negative values at similar $x$,
to allow a similar quality fit to the observables.
In this respect the flavor singlet moment $\Delta \Sigma$
is relatively stable between the different fits, with central
values ranging from 0.24 in the NNPDF14 analysis \cite{NNPDF14}
to 0.34 in the BB10 fit \cite{BB10} at $Q^2=1$~GeV$^2$.

In the higher twist sector, as indicated in Fig.~\ref{f.impact},
the twist-3 distributions $D_u$ and $D_d$ acquire unambiguous
positive and negative signs, respectively, at large $x$ values,
with magnitudes clearly different from zero.
Of most physical interest are the $x^2$-weighted moments of
$D_u$ and $D_d$, which we find to be
$\mell{D}_u(3,Q^2) =  0.013 \pm 0.005$ and
$\mell{D}_d(3,Q^2) = -0.005 \pm 0.003$ at $Q^2 = 1$~GeV$^2$.
Taking the appropriate charge squared-weighted combination of
these, one finds that for the proton the twist-3 contribution
is large, while for the neutron it mostly cancels.
This correlates with the larger higher twist effects observed for
the proton asymmetries at low $Q^2$ in Figs.~\ref{f.Apa_proton_dvcs}
and \ref{f.Apa_proton_eg1b} than in the corresponding asymmetries
for $^3$He (``neutron''), and to some extent also the deuteron.

The moments $h_p$ and $h_n$ of the twist-4 distributions are all
compatible with zero, for both the truncated and full moments.
This observation gives confidence that the \mbox{twist-3} PDFs,
and consequently the $d_2$ moments, in our analysis are determined
reliably, without significant contamination from subleading
contributions of higher twist.

\begin{figure}[t]
\centering 
\includegraphics[width=\textwidth]{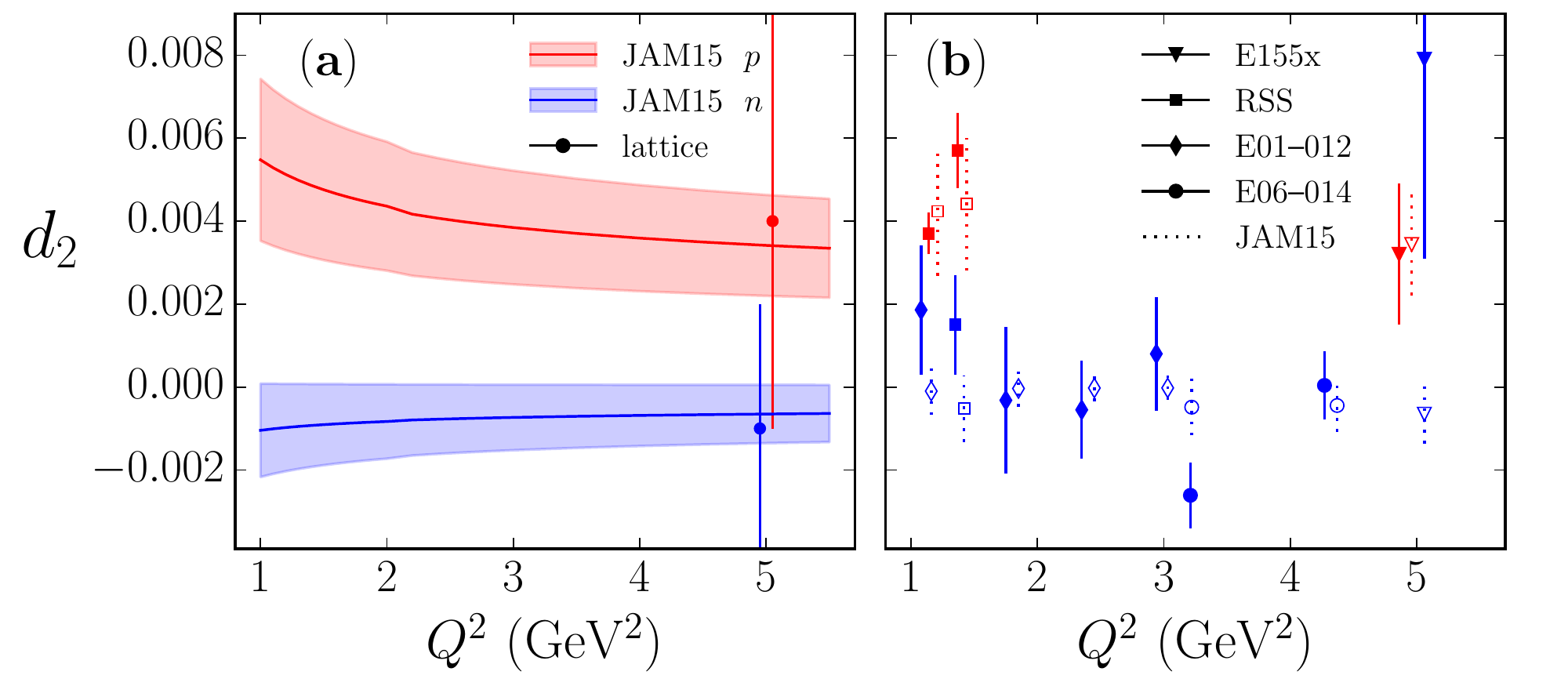}
\caption{$d_2$ moments of the proton (red curves and symbols) and
 	neutron (blue curves and symbols) computed from the JAM15
 	twist-3 $D_u$ and $D_d$ distributions and compared with
	{\bf (a)} lattice QCD calculations \cite{Goeckeler05}, and
	{\bf (b)} moments extracted from the $g_1$ and $g_2$
	structure functions from several SLAC \cite{SLAC-E155x}
	and Jefferson Lab \cite{E06-014_d2, E01-012, RSS07, RSS10}
	experiments (filled symbols), with the JAM15 results
 	(open symbols and dotted error bars) corresponding to
 	the experimentally measured regions.  The E155x results
 	include extrapolations into unmeasured regions at low and
 	high $x$, while the Jefferson Lab results are mostly from
 	the resonance region.}
\label{f.d2} 
\end{figure}

The $Q^2$ dependence of the $d_2$ moments for the proton and neutron
from the JAM15 analysis is presented in Fig.~\ref{f.d2}(a) for $Q^2$
between 1 and 5~GeV$^2$.  Note that the quoted JAM15 $d_2$ values
contain only twist-3 contributions, without TMCs [see \eref{d2tw3}],
while the corresponding experimental moments in principle contain
contributions beyond twist-3 as well as target mass effects.
For ease of notation, we will omit the explicit label
``$(\tau 3)$'' from the JAM15 $d_2$ moments in the following.
As expected from the values for the $D_u$ and $D_d$ moments
discussed above, the proton $d_2^p$ moment is positive and
\mbox{1--2~$\sigma$} away from zero, decreasing gradually from
its value $d_2^p = 0.005 \pm 0.002$ at $Q^2=1$~GeV$^2$ with
increasing $Q^2$.
The neutron $d_2^n$ moment, on the other hand, is negative
and much smaller in magnitude, but consistent with zero within
the uncertainties, $d_2^n = -0.001 \pm 0.001$.
Remarkably, without the new Jefferson Lab data, the values of
$d_2$ extracted from the global analysis (yellow curves in
Fig.~\ref{f.impact}) would be $0.005 \pm 0.002$ for the proton
and $0.005 \pm 0.005$ for the neutron.  Thus, while the proton
$d_2$ moment is essentially unchanged, the neutron central
value changes sign, although still consistent with zero.
This effect is mostly driven by the new $^3$He data from
Hall~A \cite{E06-014_d2, E06-014_A1}.
The results in Fig.~\ref{f.d2} therefore represent the most
reliable determination of the twist-3 $d_2$ moments in
global QCD analyses to date.

Our extracted $d_2$ values can also be compared with first
principles calculations of the $d_2$ matrix elements of
local twist-3 operators in lattice QCD.
In their simulations, the QCDSF/UKQCD Collaboration found
$d_2^p =  0.004(5)$ and
$d_2^n = -0.001(3)$ at a scale of $Q^2=5$~GeV$^2$
\cite{Goeckeler05}, which agrees well with the JAM15 values,
as Fig.~\ref{f.d2}(a) demonstrates.

\begin{centering}
\begin{table*}[t]
\caption{$d_2$ moments of the proton and neutron $g_{1,2}$ structure
	functions from the SLAC E155x \cite{SLAC-E155x}	and
	Jefferson Lab RSS \cite{RSS07, RSS10}, E01-012 \cite{E01-012}
	and E06-014 \cite{E06-014_d2} experiments, compared with
	the $d_2$ moments computed from the JAM15 twist-3 $D_{u,d}$
	distributions.  The $Q^2$ values and the $W$ and $x$ ranges
	for each experiment are given.  The E155x $d_2$ values
	include extrapolations into unmeasured regions, while the
	others are truncated moments over the measured regions only.
	The errors on the JAM15 values are given to the relevant
	number of significant figures, while the experimental results
	are quoted from the respective publications.\\}
\input{tab-d2}
\label{t.d2} 
\end{table*}
\end{centering}

Comparisons with $d_2$ moments extracted from the $g_1$ and $g_2$
structure functions measured in several SLAC and Jefferson Lab
experiments are illustrated in Fig.~\ref{f.d2}(b) and listed in
Table~\ref{t.d2}.  In the case of the SLAC E155x experiment,
the $d_2$ values are extrapolated from the measured region to
$x=0$ and $x=1$, while in the Jefferson Lab experiments only
the truncated moments over the measured regions are reported.
With the exception of the \mbox{E06-014} data \cite{E06-014_d2},
which partially extend into the DIS region, the truncated moments
for the Jefferson Lab experiments \cite{E01-012, RSS07, RSS10}
are restricted entirely to the nucleon resonance region.
Note that we do not include the nucleon elastic contribution
in any of the experimental or theoretical moments.
Agreement between the purely resonant empirical contributions
to $d_2$ and the twist-3 truncated moments from the JAM15 PDFs
would therefore imply the validity of quark-hadron duality for
the twist-3 spin distribution functions.  Conversely, any
differences between these may be interpreted as a violation
of duality \cite{MEK05}.

In fact, most of the experimental points for both protons and
neutrons show reasonable agreement with the JAM15 $d_2$ values
within the experimental and PDF errors.  An exception is the
lower-$Q^2$ point from Jefferson Lab E06-014, which is about
2$\sigma$ lower than the JAM15 result, and the SLAC E155x
neutron value at $Q^2=5$~GeV$^2$, which is significantly higher
(albeit with sizable uncertainty) than any of the other neutron
$d_2$ results at lower $Q^2$ and the JAM15 fit.
Future data from Jefferson Lab at 12~GeV \cite{12GeV_d2}
may enable the neutron $d_2$ moment to be determined more
precisely up to $Q^2 \approx 6$~GeV$^2$.

\section{Conclusion}
\label{s.conclusion}

We have performed a new global QCD analysis of spin-dependent
parton distributions including all available inclusive DIS data
on longitudinal and transverse polarization asymmetries from
experiments at CERN, SLAC and DESY, and new high-precision
measurements from Jefferson Lab.  The analysis is the first
performed using a newly developed fitting strategy based on
data resampling and cross validation, the key feature of which
is the iterative methodology.  This approach is fundamentally
data driven, with the prior parameters that are initially
distributed from flat sampling across parameter space iteratively
transformed into posteriors that are distributed consistently
with the information contained in the data and its uncertainties.

One of the main advantages of the iterative Monte Carlo approach
is that by sampling over a large parameter space one can avoid
introducing biases that are inherent in standard single-fit
analyses that assume a specific set of initial fitting parameters.
Since the $\chi^2$ is a highly nonlinear function of the fit
parameters, in the presence of multiple solutions any single fit
can be stuck in a local minimum and yield unreliable results for
the PDFs.  This is particularly relevant for the higher twist
distributions, for which there is considerably less experience
in global fitting.  Furthermore, being based on statistical
error analysis, the IMC procedure allows for the unambiguous
determination of PDF errors, without the need for introducing
any tolerance criteria when handling numerous data sets.

Our aim has been to maximally utilize the available data over the
greatest range of kinematics which the theoretical perturbative QCD
description permits.  To this end we evaluate both the longitudinal
and transverse asymmetries consistently up to ${\cal O}(1/Q^2)$
corrections, which necessitates including twist-3 and twist-4
contributions to the $g_1$ structure function and twist-3
corrections to $g_2$, as well as the known target mass corrections
to the leading twist and twist-3 terms.
In addition, we account for nuclear smearing effects, including
finite-$Q^2$ corrections to these, for data on deuterium and $^3$He
targets, which constitutes about 1/3 of the total database.
To empirically determine the optimal kinematic range over which the
data can be reliably fitted, we studied the sensitivity of the results
to the choice of cuts on $W^2$ and $Q^2$.  By examining the stability
of the moments of the extracted PDFs with respect to the cuts,
we could ascertain that the limits $W_{\rm cut}^2 = 4$~GeV$^2$
and $Q_{\rm cut}^2 = 1$~GeV$^2$ correspond to the boundary of the
applicability of the current global analysis.

Overall a very good description of the global inclusive DIS data set
has been obtained in our fit, over the entire range of $Q^2$ and $x$
covered by the preferred cuts.  Of the approximately 2500 data points
in the global data set, around 1400 have been added with the inclusion
of the new high-precision Jefferson Lab data, especially at lower
$Q^2$ and $W^2$.  The impact of the new data has been a general
reduction of the uncertainties on the leading twist and higher twist
distributions in the measured region.

For the $\Delta u^+$ and $\Delta d^+$ distributions, the new PDFs
are qualitatively similar to those found in previous global analyses,
with $\Delta u^+$ slightly higher at intermediate $x$ values, while
$\Delta d^+$ is somewhat less negative at large $x \gtrsim 0.1$ than
in the previous JAM13 fit \cite{JAM13}.
One of the limitations of the inclusive DIS-only analysis is the
introduction of large correlations between the nonstrange and
strange quark PDFs, which results in a slightly harder $\Delta s^+$
distribution, but one which has a clear negative sign.
Furthermore, with the addition of the lower-$Q^2$ Jefferson Lab data,
the gluon distribution, which is constrained here mainly through
$Q^2$ evolution, becomes positive across all $x$ values, and is
remarkably similar to the latest fit from Ref.~\cite{DSSV14} that
includes the recent RHIC jet data.

The biggest impact of the Jefferson Lab data, however, is in the
higher twist sector, where the new high-precision asymmetries on the
proton and deuteron from CLAS in Hall~B \cite{eg1b-p, eg1-dvcs, eg1b-d} 
and on $^3$He from Hall~A \cite{E06-014_A1, E06-014_d2} allow the
flavor dependence of the twist-3 distributions $D_u$ and $D_d$ to be
determined.  In particular, we find that the sign of the $D_d$ PDF
changes from positive to negative, which directly impacts the
determination of the twist-3 $d_2$ moments of the neutron.
Thus while the proton $d_2^p$ moment remains large (on the scale
of previous measurements) and positive, the new neutron $d_2^n$
moment becomes negative, although still compatible with zero to
within 1$\sigma$.  Interestingly, the JAM15 $d_2$ results agree
well with the available lattice QCD calculations at $Q^2=5$~GeV$^2$
\cite{Goeckeler05} for both the proton and neutron, but disagree
with the magnitude and sign of the neutron $d_2^n$ moment extracted
from the SLAC E155x experiment \cite{SLAC-E155x}.

In the future, data from 12~GeV Jefferson Lab experiments will allow
the $d_2$ moments to be determined more precisely in the DIS region
at higher $Q^2$ values \cite{12GeV_d2}, and also provide stronger
constraints on the large-$x$ behavior of PDFs through precise
measurements of polarization asymmetries over a greater range of
$Q^2$ and $W^2$ \cite{12GeV_HallC, 12GeV_HallA}.
In the shorter term, the current analysis will be extended to include
semi-inclusive DIS asymmetries, which will place stronger constraints
on the sea quark polarization, as well as jet and $\pi$ production
asymmetries in polarized $pp$ collisions \cite{JAM16}.
In view of the importance of determining the proton spin decomposition
into its constituent components, it will be of great interest to
explore the emergent picture for the sea quark and gluon polarization
within the IMC approach.

\section*{Acknowledgments}

We are grateful to M.~Stratmann and W.~Vogelsang for assistance
with Mellin moment techniques, and to C.~Fern\'{a}ndez-Ram\'{i}rez,
P.~Jimenez-Delgado, F.~M.~Steffens and the experimental members
of the JAM Collaboration \cite{JAMweb} for helpful discussions.
This work was supported by the US Department of Energy (DOE) contract
No.~DE-AC05-06OR23177, under which Jefferson Science Associates, LLC
operates Jefferson Lab.
A.A. was partially supported by the DOE contract No.~DE-SC008791,
S.K. was supported by the DOE under contract DE-FG02-96ER40960,
and N.S. was partially supported by GAUSTEQ (Germany and U.S. Nuclear
Theory Exchange Program for QCD Studies of Hadrons and Nuclei),
DOE contract No.~DE-SC0006758.

\newpage
\appendix
\section{Notations}

In this appendix we provide for convenience a summary of the
notations used in this work for several common moments of
twist-2, twist-3 and twist-4 distributions.  In general,
we define the $N$-th Mellin moment of a function $f(x)$ by
\begin{align}
\mell{f}(N,Q^2)
&= \int_0^1 dx\, x^{N-1}\, f(x,Q^2),
\label{e.fN}
\end{align}
which is a continuous functions of $N$.  To distinguish the
moments $\mell{f}(N,Q^2)$ from the $x$-dependent distributions
$f(x,Q^2)$, we denote these in boldface.  Table~\ref{t.notation}
summarizes the different notations used according to
Eq.~(\ref{e.fN}) here and elsewhere in the literature.

\begin{centering}
\begin{table*}[ht]
\caption{Summary of notations used in this work for some moments
	of twist-2, twist-3 and twist-4 distributions, including
	the formal notation as defined in Eq.~(\ref{e.fN}) and the
	definitions in terms of integrals of PDFs and structure
	functions. \\}
\begin{tabular*}{0.9\textwidth}{@{\extracolsep{\fill}} lll} \hline\hline
\multicolumn{1}{l}{shorthand}
& \multicolumn{1}{l}{formal}
& \multicolumn{1}{l}{definition}		\\ \hline
%
%
%
~$\Delta \Sigma(Q^2)$ &
$\sum_q\mell{\Delta} \mell{q}^+(1,Q^2)$ &
$\sum_q\int_0^1dx~\Delta q^+(x,Q^2)$ \\
~$\Delta G(Q^2)$ &
$\mell{\Delta} \mell{g}(1,Q^2)$ &
$\int_0^1dx~\Delta g(x,Q^2)$ \\
~$d_2(Q^2)$ &
$2\mell{g}_1(3,Q^2)+3\mell{g}_2(3,Q^2)$ &
$\int_0^1 dx\, x^2\, [ 2g_1(x,Q^2)+3g_2(x,Q^2) ]$ \\
~$d_2^{(\tau 3)}(Q^2)$ &
$2\mell{g}_1^{(\tau 3)}(3,Q^2)+3\mell{g}_2^{(\tau 3)}(3,Q^2)$ &
$\int_0^1dx\, x^2\, [ 2g_1^{(\tau 3)}(x,Q^2)+3g_2^{(\tau 3)}(x,Q^2) ]$ \\
  &
$= \sum_q~e_q^2~\mell{D}_q(3,Q^2)$ &  \\
~$h(Q^2)$ &
$\mell{H}(3,Q^2)$ &
$\int_0^1 dx~x^2~H(x,Q^2)$ \\ \hline
\label{t.notation}
\end{tabular*}
\end{table*}
\end{centering}
\clearpage


\end{document}

%% file: tab-chi2.tex
\scriptsize
\begin{tabular}{lccccl} 			\hline\hline
experiment &~reference~ &~observable~   &~target~    &~\# points~~  & $\chi^2_{\rm dof}$~~ \\ \hline
EMC
 & \cite{EMC89}		& $A_1$		& $p$        & 10          & 0.40 \\
SMC
 & \cite{SMC98}		& $A_1$		& $p$        & 12          & 0.47 \\
SMC
 & \cite{SMC98}		& $A_1$		& $d$        & 12          & 1.62 \\
SMC
 & \cite{SMC99}		& $A_1$		& $p$        & 8           & 1.26 \\
SMC
 & \cite{SMC99} 	& $A_1$		& $d$        & 8           & 0.57 \\
COMPASS
 & \cite{COMPASS10}	& $A_1$		& $p$        & 15          & 0.92 \\
COMPASS
 & \cite{COMPASS07}	& $A_1$		& $d$        & 15          & 0.67 \\
COMPASS
 & \cite{COMPASS16}	& $A_1$		& $p$        & 51          & 0.76 \\
SLAC E80/E130~~
 & \cite{SLAC-E130}	& $A_\parallel$ & $p$	     & 22          & 0.59 \\
SLAC E142
 & \cite{SLAC-E142}	& $A_1$		& $^3\rm He$ & 8           & 0.49 \\
SLAC E142
 & \cite{SLAC-E142}	& $A_2$		& $^3\rm He$ & 8           & 0.60 \\
SLAC E143
 & \cite{SLAC-E143}	& $A_\parallel$ & $p$        & 81          & 0.80 \\
SLAC E143
 & \cite{SLAC-E143}	& $A_\parallel$	& $d$        & 81          & 1.12 \\
SLAC E143
 & \cite{SLAC-E143} 	& $A_\perp$	& $p$        & 48          & 0.89 \\
SLAC E143
 & \cite{SLAC-E143} 	& $A_\perp$	& $d$        & 48          & 0.91 \\
SLAC E154
 & \cite{SLAC-E154} 	& $A_\parallel$	& $^3\rm He$ & 18          & 0.51 \\
SLAC E154
 & \cite{SLAC-E154} 	& $A_\perp$	& $^3\rm He$ & 18          & 0.97 \\
SLAC E155
 & \cite{SLAC-E155p}	& $A_\parallel$	& $p$        & 71          & 1.20 \\
SLAC E155
 & \cite{SLAC-E155d} 	& $A_\parallel$	& $d$        & 71          & 1.05 \\
SLAC E155
 & \cite{SLAC-E155_A2pd} & $A_\perp$	& $p$        & 65          & 0.99 \\
SLAC E155
 & \cite{SLAC-E155_A2pd} & $A_\perp$	& $d$        & 65          & 1.52 \\
SLAC E155x
 & \cite{SLAC-E155x} & $\widetilde{A}_\perp$ & $p$   & 116         & 1.27 \\
SLAC E155x
 & \cite{SLAC-E155x} & $\widetilde{A}_\perp$ & $d$   & 115         & 0.83 \\
HERMES
 & \cite{HERMES97}	& $A_1$		& ``$n$''    & 9           & 0.25 \\
HERMES
 & \cite{HERMES07}	& $A_\parallel$	& $p$        & 35          & 0.47 \\
HERMES
 & \cite{HERMES07} 	& $A_\parallel$	& $d$        & 35          & 0.94 \\
HERMES
 & \cite{HERMES12}	& $A_2$		& $p$        & 19          & 0.93 \\
JLab E99-117
 & \cite{E99-117}	& $A_\parallel$	& $^3\rm He$ & 3           & 0.27 \\
JLab E99-117
 & \cite{E99-117}	& $A_\perp$	& $^3\rm He$ & 3           & 1.58 \\
JLab E06-014
 & \cite{E06-014_A1}	& $A_\parallel$	& $^3\rm He$ & 14          & 2.12 \\
JLab E06-014
 & \cite{E06-014_d2}	& $A_\perp$	& $^3\rm He$ & 14          & 1.06 \\
JLab eg1-dvcs
 & \cite{eg1-dvcs}	& $A_\parallel$	& $p$        & 195         & 1.52 \\
JLab eg1-dvcs
 & \cite{eg1-dvcs}	& $A_\parallel$	& $d$        & 114         & 0.94 \\
JLab eg1b
 & \cite{eg1b-p}	& $A_\parallel$	& $p$        & 890         & 1.11 \\
JLab eg1b
 & \cite{eg1b-d}	& $A_\parallel$	& $d$        & 218         & 1.02 \\ \hline
%
total &       &                   &            & 2515        & 1.07 \\
\hline
\end{tabular}
\normalsize

%% file: tab-d2.tex
\begin{tabular*}{0.9\textwidth}{@{\extracolsep{\fill}} lcccccrl}		\hline\hline
experiment & ref. &  target  & $Q^2$   & $W$ range    & $x$ range
 & \multicolumn{1}{c}{~$d_2(\rm JAM15)$}
 & \multicolumn{1}{c}{$d_2(\rm exp.)$~~}				\\
 & &                              &(GeV$^2$)& (GeV)        & & & 	\\ \hline
E155x   & \cite{SLAC-E155x} & $p$ &  5.00   & $> M$        & [0, 1]       & $0.003(1)$~  & $ ~~0.0032(17)$ \\
        & \cite{SLAC-E155x} & $n$ &  5.00   & $> M$        & [0, 1]       & $-0.0007(7)$ & $ ~~0.0079(48)$ \\
RSS     & \cite{RSS07}      & $p$ &  1.30   & [1.06, 2.01] & [0.29, 0.84] & $0.004(2)$~  & $ ~~0.0057(9)$ \\
        & \cite{RSS10}      & $p$ &  1.28   & [1.08, 1.91] & [0.32, 0.82] & $0.004(2)$~  & $ ~~0.0037(5)$ \\
        & \cite{RSS10}      & $n$ &  1.28   & [1.08, 1.91] & [0.32, 0.82] & $-0.0005(8)$ & $ ~~0.0015(12)$ \\
E01-012 & \cite{E01-012}    & $n$ &  1.20   & [1.04, 1.38] & [0.54, 0.86] & $-0.0001(6)$ & $ ~~0.00186(156)$ \\
        & \cite{E01-012}    & $n$ &  1.80   & [1.09, 1.56] & [0.54, 0.86] & ~$0.0000(4)$ & $-0.00032(177)$ \\
        & \cite{E01-012}    & $n$ &  2.40   & [1.07, 1.50] & [0.64, 0.90] & ~$0.0000(3)$ & $-0.00055(118)$ \\
        & \cite{E01-012}    & $n$ &  3.00   & [1.10, 1.61] & [0.64, 0.90] & ~$0.0000(3)$ & $ ~~0.00080(137)$ \\
E06-014 & \cite{E06-014_d2} & $n$ &  3.21   & [1.11, 3.24] & [0.25, 0.90] & $-0.0005(7)$ & $-0.00261(79)$ \\
        & \cite{E06-014_d2} & $n$ &  4.32   & [1.17, 3.72] & [0.25, 0.90] & $-0.0005(6)$ & $ ~~0.00004(83)$ \\ \hline
\end{tabular*}